\documentclass[10pt,twocolumn,twoside]{IEEEtran}
\makeatletter
\def\ps@headings{%
\def\@oddhead{\mbox{}\scriptsize\rightmark \hfil \thepage}%
\def\@evenhead{\scriptsize\thepage \hfil \leftmark\mbox{}}%
\def\@oddfoot{}%
\def\@evenfoot{}}
\makeatother

\ifCLASSINFOpdf
\else
\fi

\usepackage{url}
\usepackage{amssymb}
\usepackage{amsfonts}
\usepackage{amsmath}
\usepackage{amsmath,amssymb,verbatim}
\usepackage{graphicx}
\usepackage{float}
\usepackage[lined,linesnumbered,ruled,vlined]{algorithm2e}
\usepackage{epsfig}
\usepackage{subfig}%
\usepackage{url}
\usepackage{color}
\usepackage{delarray}

\usepackage{amsthm}
\usepackage{comment}
\usepackage{ifthen}

\newboolean{Longversion} \setboolean{Longversion}{true}

\newtheorem{proposition}{Proposition}
\newtheorem{theorem}{Theorem}
\newtheorem{definition}{Definition}
\newtheorem{corollary}{Corollary}

\begin{document}
\title{Decentralized Protection Strategies \\against SIS Epidemics in Networks}
\author{Stojan Trajanovski,~\IEEEmembership{Student Member,~IEEE,}
        Yezekael Hayel,~\IEEEmembership{Member,~IEEE,}
       Eitan Altman,~\IEEEmembership{Fellow,~IEEE,}\\
       Huijuan Wang,
        and~Piet Van Mieghem,~\IEEEmembership{Member,~IEEE}% <-this % stops a space
\thanks{The first two authors have contributed equally.}
\thanks{S. Trajanovski,  H. Wang and P. Van Mieghem are with Delft University of Technology, Delft,
The Netherlands (e-mails:  \{S.Trajanovski, H.Wang, P.F.A.VanMieghem\}@tudelft.nl).}% <-this % stops a space
\thanks{Y. Hayel is with the University of Avignon, Avignon, France (yezekael.hayel@univ-avignon.fr).}% <-this % stops a space
\thanks{E. Altman is with INRIA, Sophia Antipolis, France (eitan.altman@inria.fr).}%
%\thanks{Manuscript received April 19, 2005; revised December 27, 2012.}
}

\date{}
\maketitle

%\author{
%\IEEEauthorblockN{Eitan Altman}
%\IEEEauthorblockA{INRIA\\
%Sophia Antipolis, France\\
%eitan.altman@sophia.inria.fr}
%\and
%\IEEEauthorblockN{Yezekael Hayel}
%\IEEEauthorblockA{University of Avignon\\
%Avignon, France\\
%yezekael.hayel@univ-avignon.fr}
%\and
%\IEEEauthorblockN{Stojan Trajanovski, Huijuan Wang, Piet Van Mieghem}
%\IEEEauthorblockA{Delft University of Technology\\
%Delft, The Netherlands\\
% \{S.Trajanovski, H.Wang, P.F.A.VanMieghem\}@tudelft.nl}
%}

\begin{abstract}
Defining an optimal protection strategy against viruses, spam propagation or any other kind of contamination process is an important feature for designing new networks and architectures. In this work, we consider decentralized optimal protection strategies when a virus is propagating over a network through a SIS epidemic process. We assume that each node in the network can fully protect itself from infection at a constant cost, or the node can use recovery software, once it is infected.

We model our system using a game theoretic framework and find pure, mixed equilibria, and the Price of Anarchy (PoA) in several network topologies. Further, we propose both a decentralized algorithm and an iterative procedure to compute a pure equilibrium in the general case of a multiple communities network. Finally, we evaluate the algorithms and give numerical illustrations of all our results.
\end{abstract}


%\linespread{0.99}
%\addtolength{\floatsep}{-0.1mm}
%\addtolength{\dblfloatsep}{-0.1mm}

\section{Introduction}\label{sec:Introduction}
% Motivation viruspread & immunization
Virus spread processes in networks can be explained, using epidemic models~\cite{Omic09,Chakrabarti2008,Ganesh2005,Kephart91,PhysRevE.88.022801}. The probability of infection, especially in the steady-state, in relation to the properties of the underlying network has been widely studied in the past~\cite{Omic09, Chakrabarti2008}. We consider the Susceptible Infected Susceptible (SIS) model, which is one of the most studied epidemic models~\cite{Omic09,VanMieghem2012}. In the SIS model, the state of each node is either susceptible or infected. The recovery (curing) process of each infected node is an independent Poisson process with a recovery rate $\delta$. Each infected node infects each of its susceptible neighbors with a rate $\beta$, which is also an independent Poisson process. Immunization~\cite{Cohen2003,Schneider2011} (e.g. via antivirus software) or quarantining~\cite{Omic2010} (via modular partitioning) fully prevent nodes from being infected, while additional tools, like anti-spyware software, can clean the virus from an infected node. Finally, the network can be modified to increase the epidemic threshold~\cite{PhysRevE.84.016101}.

% Our problem
This paper considers investment games that find appropriate protection strategies against SIS virus spread. In particular, we consider a game, in which, each node is a player in the game and decides individually whether or not to invest in antivirus protection. Further, if a host does not invest in antivirus protection, it remains vulnerable to the virus spread process, but can recover (e.g., by a system recovery or clean-up software). The cost or negative utility of each node (player) is: (i) the investment cost, if the node decides to invest in antivirus software or in the opposite case: (ii) the cost of being infected, which is proportional to the infection probability in the epidemic steady-state.

Models in game theory describe complex systems, in which several decision makers optimize their own objective and interact together. In our setting, the decision of each node has a natural impact on the cost of the other players through the network structure and the epidemic dynamics. In fact, a node which is not protected would be a potential relay of the virus. Moreover, a node protecting itself additionally induces a protection to its neighbors. This last concept is known as a ``positive externality effect'' in economic game theory~\cite{fudenberg1991game}. Finally, as the network under study may be potentially large, we look for a scalable optimal strategy, which is given by the equilibrium of a game. Thus, a game theory model is suitable to study such decentralized protection strategy problem in a network.

% Contribution
Our main contributions are summarized as follows:
\begin{enumerate}
\item We prove that the game on a \emph{single community/full mesh} network is a potential game by showing that it is equivalent to a congestion game. Subsequently, we determine a closed-form expression for the unique pure equilibrium. We also prove the existence and uniqueness of a mixed equilibrium.
\item We provide a measure of the equilibrium efficiency based on the Price of Anarchy (PoA) and propose a simple, fully decentralized Reinforcement Learning Algorithm (RLA) that converges to the equilibrium.
\item We extend our equilibrium analysis to \emph{bipartite} networks, where we show that multiple equilibria are possible. At an equilibrium, the number of nodes that invest in one partition is often close to the number of nodes that invest in the other partition.  
\item In a \emph{multi-communities} network, in which several communities are communicating through a single core node, we introduce the concept of parametric potential games, and subsequently, show the convergence of an iterative procedure to a pure Nash equilibrium. Here, we also show that the iterative procedure can be used to find the metastable stationary infection probability of the core node that links all the communities.
%\item For the \emph{community network}, in the framework of parametric potential games, we find ...
%We employ the game theoretic setting into a game with multiple communities with different sizes
\end{enumerate}

% Outline
The paper is organized as follows. An overview of decentralized protection strategies, epidemics and game theory is given in Section~\ref{sec:relatedWork}. The SIS epidemic model is introduced in Section~\ref{sec:SISEpidemics}. Sections~\ref{sec:GameModelComplete} and \ref{sec:BipartiteNetwork} describe the game models in a single community (full mesh) and bipartite network, respectively and subsequently prove game theoretic results. In Section~\ref{sec:CommunityNetwork}, we study the potential parametric games and equilibrium properties in a \emph{multi-communities} network. Evaluation of the RLA algorithm, its convergence to the equilibrium point and numerical illustrations are given in Section~\ref{sec:Simulations}. We conclude in Section~\ref{sec:Conclusion}. The proofs of the propositions and corollaries are given in Appendices~\ref{ProofsCompleteGraph}, \ref{ProofsBiPartite} and \ref{ProofsMultiComm}.%An overview of the related work is given in Section~\ref{sec:relatedWork}. 

\section{Related work}\label{sec:relatedWork}
% Epidemic spread in networks
Virus spread processes in networks have been studied in the past~\cite{Omic09,Chakrabarti2008,Ganesh2005,Kephart91}, usually considering the number of infected nodes~\cite{Omic09} over the time and in stationary regimes, the epidemic threshold~\cite{Chakrabarti2008} or the relation with eigenvalues~\cite{VanMieghem2012EPL}. One of the widely explored Susceptible Infected Susceptible (SIS) approximations is the N-intertwined mean-field approximation NIMFA~\cite{Omic09,VanMieghem2012}.

Game theoretical studies for network problems have been conducted, in routing~\cite{Orda1993,Altman2000}, network flow~\cite{Korilis1995}, workload on the cloud~\cite{Nahir2012} or optimal network design~\cite{Nahir2013,Gourdin2011}, employing standard game-theoretic concepts~\cite{Koutsoupias2009,Roughgarden2002} such as pure Nash or mixed equilibrium. The Price of Anarchy (PoA)~\cite{NisanRoughgardenTardosVazirani(AlgoGameTheory)07,Roughgarden2002} is often used as an equilibrium performance evaluation metric.

% Game theory in networks and epidemics and differences from our work
Game theory has been  used in several studies~\cite{5062065,Aspnes2006,TechReport_Acemoglu2013,Jiang2008,Lelarge2008,Kearns04algorithmsfor,PreciadoCDC2013,PreciadoZS13,Theodorakopoulos2013,AAAI148478} related to epidemic protection or curing, for example, in a generalized game settings~\cite{Lelarge2008} without considering the infection state of the neighbors; by assigning nodal weights to reflect the security level~\cite{Jiang2008} etc. Omi\'{c} \emph{et al.}~\cite{5062065} tune the strength of the nodal antivirus protection i.e. how big those (different) $\delta$ should be taken. Contrarily to~\cite{5062065}, (i) we fix the curing and infection rates, which are not part of the game, and the decision consists of a player's choice to invest in an antivirus or not; (ii) we also consider mixed strategies Nash Equilibrium and (iii) propose a convergence algorithm to the equilibrium point. The goal of~\cite{5062065} is in finding the optimal curing rates $\delta_{i}$ for each player $i$, while this paper targets the optimal decision of taking an anti-virus that fully protects the host, because today's antivirus software packages provide accurate and up-to-date virus protection.% The related papers on security games~\cite{Aspnes2006,TechReport_Acemoglu2013,Jiang2008,Lelarge2008,Kearns04algorithmsfor} are usually applied in non-SIS environments (e.g., (i) without considering the infection state of the neighbors and (ii) without an additional mechanism for recovery), for generalized game settings~\cite{Lelarge2008} or by assigning nodal weights to reflect the security level~\cite{Jiang2008}. % Aspnes2007

\section{SIS epidemics on networks}\label{sec:SISEpidemics}
We consider a connected, undirected and unweighted network $G$ with $N$ nodes. The virus behaves as an SIS epidemic, where an infected node can infect each of its direct, healthy neighbors with rate $\beta$. Each node can be cured at rate $\delta$, after which the node becomes healthy, but susceptible again to the virus. Both infection and curing
process are independent Poisson processes. All nodes in the network $G$  are prone to a virus that can re-infect the nodes multiple times.

We denote the viral probability of infection for node $i$ at time $t$ by $v_{i}\left(N; t\right)$. For each node $i$ of the graph with $N$ nodes, the SIS governing equation, under the standard $N$-Intertwined mean-field approximation (NIMFA)~\cite{VanMieghem2011}, is given by%
\small{
\begin{align}
\frac{dv_{i} \left(N; t\right) }{d t}=-\delta v_{i} \left(N; t\right)
+\beta (1-v_{i} (N;t))\sum\limits_{j=1}^{N} a_{ij} v_{j}\left(N; t\right), \label{governEq:SIS}
\end{align}
}\normalsize where $a_{ij}= 1$, if nodes $i$ and $j$ are directly connected by a link, otherwise $a_{ij}=0$. The physical interpretation of (\ref{governEq:SIS}) is the following: the infection probability of a node $i$ changes over time by two competing processes: (i) each infected neighbor of node $i$ tries to infect him with Poisson rate $\beta$,  while  node $i$ is healthy (with probability $1 - v_{i} (t)$), and (ii) node $i$ can be cured with a Poisson rate $\delta$, while infected with probability $v_{i} (t)$.

We further confine ourselves to the stationary regime of the SIS epidemic process, meaning $\lim \limits_{t\rightarrow\infty}\frac{dv_{i} (N;t)}{dt}=0$. We
denote the spreading rate $\tau=\frac{\beta}{\delta}$ and $v_{i,\infty} (N)=\lim \limits_{t\rightarrow\infty}v_{i}\left(N; t\right) $ the probability of node $i$ being infected in the stationary regime. In the stationary regime, based on (\ref{governEq:SIS}),
%\small{
%\begin{align}
%0=-v_{i,\infty} \left(N\right)
%+\tau (1-v_{i,\infty} (N))\sum\limits_{j=1}^{N} a_{ij} v_{j,\infty }\left(N\right) \label{governEq:SISStat}
%\end{align}
%}\normalsize 
$v_{i,\infty} \left(N\right)$ can be expressed as~\cite{Omic09,VanMieghem2011},
\small{
\begin{align}
v_{i,\infty} (N)&=1-\frac{1}{1+\tau\sum\limits_{j=1}^{N} a_{ij} v_{j,\infty} (N)}
\label{steady_state_NIMFA_node_i}
\end{align}
}\normalsize for $\forall i=1,2,\ldots,N$. These steady-state equations only have two possible solutions: (i) the trivial $v_{i\infty} (N)=0$, corresponding to the exact absorbing state in SIS epidemics, and (ii) the non-trivial solution ($v_{i\infty} (N)\neq 0$), corresponding to the metastable SIS regime. In this paper, we focus on the metastable SIS regime.
%
%%scaling the time as $t^{\ast}=\delta t$, is given by
%\small{
%\begin{align*}
%\frac{dv_{i} \left(N; t\right) }{dt^{\ast}}=-v_{i} \left(N; t\right)
%+\tau (1-v_{i} (N;t))\sum\limits_{j=1}^{N} a_{ij} v_{j}\left(N; t\right)
%\end{align*}
%}\normalsize where $\tau=\frac{\beta}{\delta}$ is the effective spreading rate. We
%further confine ourselves to the stationary regime of the SIS epidemic process, meaning $\lim_{t\rightarrow\infty}\frac{dv_{i} (N;t)}{dt}=0$, and we
%denote $v_{i\infty } (N)=\lim_{t\rightarrow\infty}v_{i}\left(N; t\right) $. The steady-state equations, valid
%for any graph~\cite{Omic09,VanMieghem2011}, are
%\small{
%\begin{align}
%v_{i\infty} (N)&=1-\frac{1}{1+\tau\sum\limits_{j=1}^{N} a_{ij} v_{j\infty} (N)}
%\label{steady_state_NIMFA_node_i}
%\end{align}
%}\normalsize for $i=1,2,\ldots,N$. These steady-state equations only have two possible solutions: the trivial $v_{i\infty} (N)=0$, corresponding to the exact absorbing state in SIS epidemics, and the non-trivial solution, corresponding to the metastable SIS regime.

The infection probabilities can be substantially different after some nodes decide to invest in protection, causing those nodes not to be part of the epidemic process. Motivated by social networks, we start with a single community network, then continue with a general multi-community network as a model of social or \emph{ego-centric social} networks~\cite{Trajanovski2013363}\footnote{An \emph{ego-centric social} network is a network of a user and its (online) friends.}. The single community game can be regarded as a simpler and special case of the multi-community social network, however, the game on a complete graph can also be applied in full mesh wireless networks (e.g., MANETs). A multi-community network can be used for modeling a network, where nodes are connected if they belong to the same community school, institution, geographical location, interest in sport or music activities. These communities may weakly overlap with a small number of nodes that belong to multiple communities. A different type of a 2-communities network is the bipartite graph. A bipartite graph models two communities, where nodes do not communicate internally in their community, and there is only inter-communities communication. Like the single community network, the bipartite network has applications, different from social networks, to reliable client-server dependences: the topology of the Amsterdam Internet Exchange is designed as a bipartite network such that all the locations in Amsterdam are connected to two high throughput Ethernet switches; the topologies of sensor networks are also bipartite graphs.
%
%
%
%
%Three scenarios are considered. The first one is a single community game, which could be regarded as a simple social network or a wireless and other full mesh networks (e.g., MANETs). We also study bipartite networks, often employed in the design of telecommunication networks. The main reason is that a bipartite topology provides satisfactory level of robustness after node or link failures. The topology of the Amsterdam Internet Exchange is designed as a bipartite network such that all the locations in Amsterdam are connected to two high throughput Ethernet switches. In addition, the topologies of sensor networks are also bipartite graphs. The third, we consider a topology inspired by online social networks, where agents are connected if they belong to the same community school, institution, geographical location, interest in sport or music activities. These communities may weakly overlap via small number of nodes that belong to multiple communities.
%
%Taking a motivation of the social networks, we first start with a single community network, then continue with specifically connected two communities network (bipartite) and finish with a general multi-community network that well-models social or ego-centric social networks. (One community network is a special case of the last one

\subsection{Single community (full mesh) network}\label{SIScompleteGraph}%: Complete Graph}
We first consider a single community (or full mesh) network, modeled as a complete graph $K_N$, where $a_{ij}=1$ for all $i$ and $j$. If some nodes are removed from the original graph, the resulting graph $K_n$ is also a complete graph, where $n \in \{0,1,\ldots, N\}$. By symmetry, all $v_{i,\infty} (n)$ are equal. For simplicity, we use the notation $v_{\infty} (n)=v_{i,\infty} (n)$. From (\ref{steady_state_NIMFA_node_i}), we have~\cite{Omic09,VanMieghem2012},
%\small{
%\begin{align*}
%v_{\infty} (N) &=1-\frac{1}{1+\tau\sum\limits_{j =1 \atop j \neq i}^{N} v_{\infty} (N)} = 1-\frac{1}{1+\tau (N-1) v_{\infty} (N)}
%\end{align*}
%}\normalsize from which we obtain
\small{
\begin{equation}
v_{\infty} (n) = v_{i,\infty} (n)=
\begin{array}\{{cl}.
1-\frac{1}{\tau (n-1)}, & \mbox{if}\quad \tau\geq\frac{1}{n-1} \text{ and }n\ge2,\\ 
0,& \mbox{otherwise}
\label{infection_prob_node_i_regular_graph}
\end{array}
\end{equation}
}\normalsize for each node $i$ in a complete graph.
% $v_{i,\infty } (N) = v_{j,\infty } (N)=v_{\infty} (N)$ for any node $i$ and $j$.

\subsection{Bipartite network}
\begin{figure}[h!tb]
\centering
\subfloat[]{
\includegraphics[trim = 0mm 0mm 10mm 0mm,clip,width=0.205\textwidth]{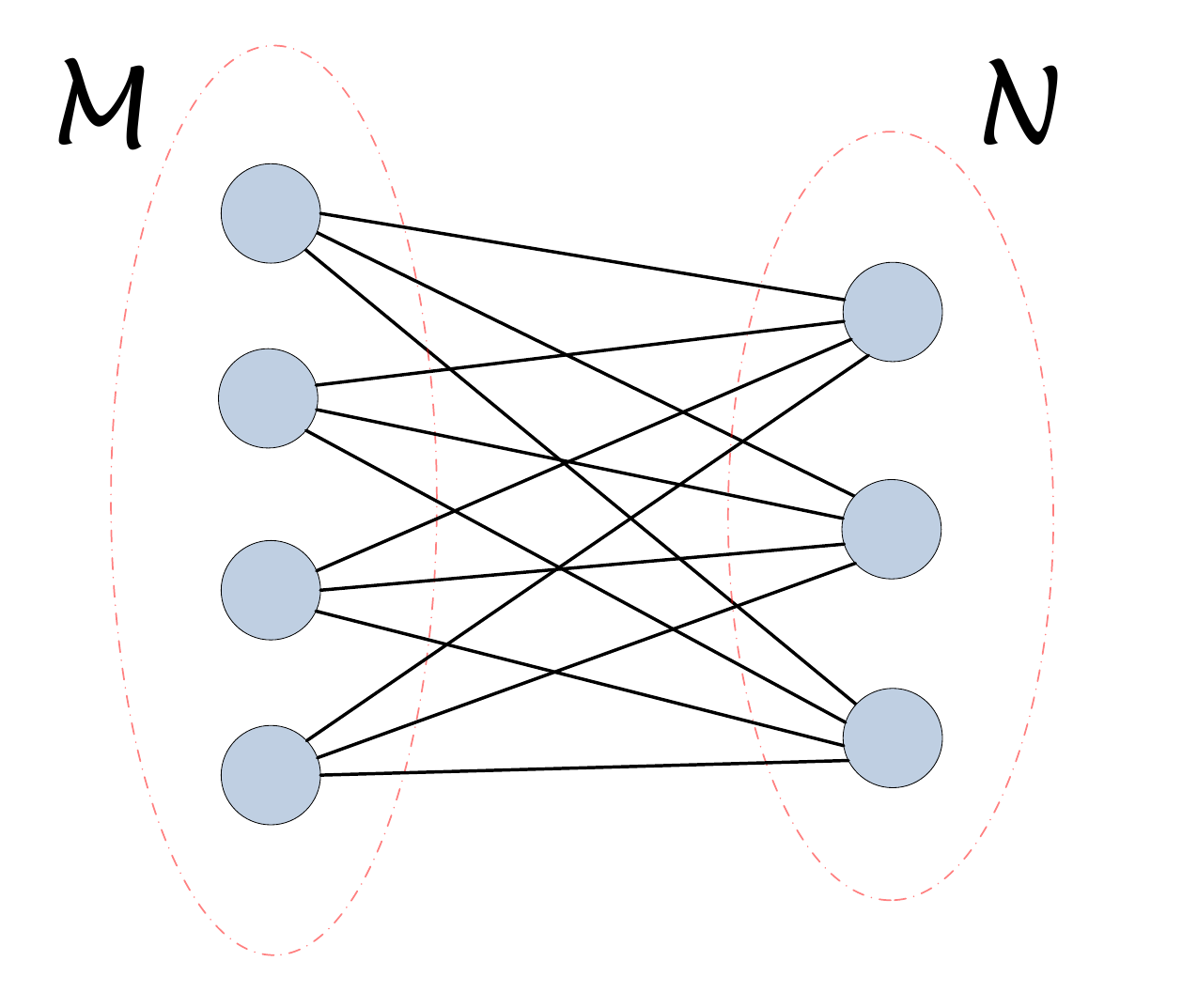}
\label{fig:bipartite}}
\subfloat[]{
\includegraphics[trim = 5mm 17mm 0mm 10mm,clip,width=0.275\textwidth]{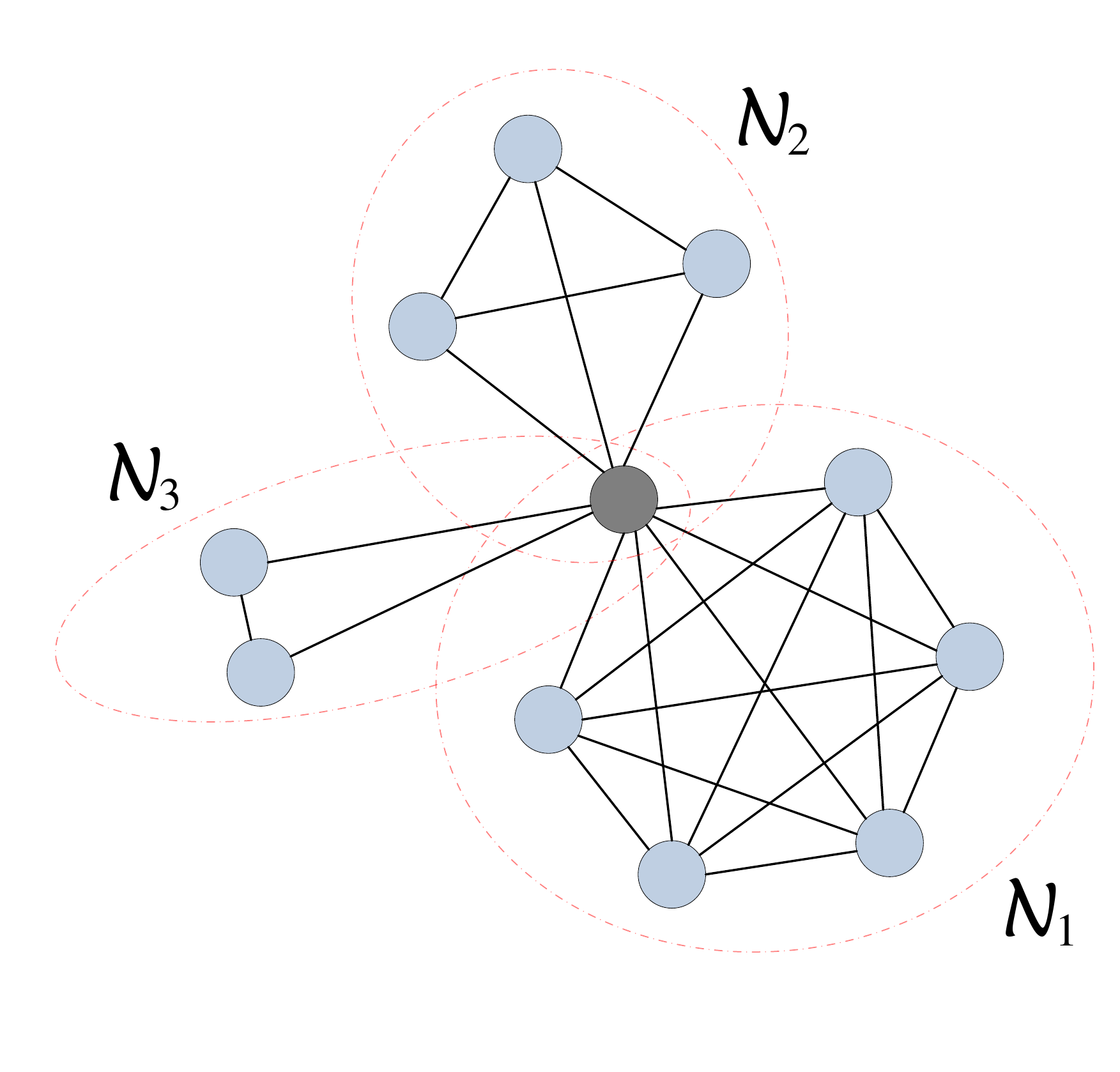}
\label{fig:communityGraph}}

\caption[]{(a) An example of a bipartite graph and (b) an example of a multi-communities network.} \label{fig:regRunnTime}
\end{figure}

The bipartite network $K_{M,N}$ consists of two clusters, $\mathcal{M}$ and $\mathcal{N}$, with $M$ and $N$ nodes, such that a node from a cluster is connected to all the nodes from the other cluster and is not connected to any node within his own cluster as illustrated in Fig.~\ref{fig:bipartite}. Therefore, there are exactly $(M+N)$ nodes and $MN$ links in the network. 
\vspace{-0.5cm}

%
%\begin{figure}[tbp]
%\begin{center}
%\includegraphics[trim = 0mm 0mm 0mm 0mm,clip,width=0.28\textwidth]{images/bipartiteGraph}
%\end{center}
%\caption{Example of a bipartite graph.}
%\label{fig:bipartite}
%\end{figure}
If some nodes are removed from the original graph, the resulting graph is again a bipartite graph $K_{m,n}$, where $m$ and $n$ are the number of the remaining nodes in the clusters $\mathcal{M}$ and $\mathcal{N}$, respectively.

The system of governing equations (\ref{steady_state_NIMFA_node_i}), for $K_{m,n}$, reduces to~\cite{Omic09,4610080}
\small{
\begin{align}
v_{\infty}^{(\mathcal{M})} (m,n) = \frac{\tau^2 mn - 1}{\tau m (\tau n + 1)} \text{  and   } v_{\infty}^{(\mathcal{N})} (m,n) = \frac{\tau^2 mn - 1}{\tau n (\tau m + 1)}
\label{solutionsBipartite}
\end{align}
}\normalsize where $v_{\infty}^{(\mathcal{M})} (m,n)$ and $v_{\infty}^{(\mathcal{N})} (m,n)$ are the infection probabilities in clusters $\mathcal{M}$ and $\mathcal{N}$, respectively.

\subsection{Multi-communities network}
We consider the \emph{multi-communities network} composed of $M$ cliques/communities $\{\mathcal{N}_m \mid m =1,2,\ldots,M\}$, where each community has size $|\mathcal{N}_m | = N_m+1$. Furthermore, we assume that each sub-network is connected to any other sub-network by a \emph{core node}, and all communities overlap (communicate) only via one node as shown in Fig.~\ref{fig:communityGraph}. The core node does not participate
into the game and cannot invest in an anti-virus protection. The total number of nodes in this network is $1+\sum_{m=1}^{M}N_m$. This communities network is highly modular~\cite{TrajanovskiEpjB2012} and has the property that after removing some nodes from arbitrary communities, the resulting graph is still a communities network with communities sizes $(n_m +1)$ for all $m$. The degree of any node $i$, which belongs to community $%
\mathcal{N}_m $, but is not the core node, is equal to $n_m$.% (all the other nodes from his community plus the
%core node).%An example of such topology is depicted in Fig.~\ref{fig:communityGraph}. In fact, this node permits to keep connexions between subgraphs (communities).
%
%\begin{figure}[h!tb]
%\begin{center}
%\includegraphics[trim = 0mm 17mm 0mm 10mm,clip,width=0.34\textwidth]{images/communityGraph}
%\end{center}
%\caption{Example of a multi-communities network. The core node is given in dark gray color.}
%\label{fig:communityGraph}
%\end{figure}
%  for all $m\in \{1,\ldots,M\}$

The core node functions as a bridge between all communities: the $M$ fully connected
graphs are interconnected through one common node. We will compute the infection probability for each non-core node in each community as a function of the infection probability of a core node. Different communities might have different virus-spread dynamics. We consider the more general in-homogeneous SIS model, see e.g.,~\cite{VanMieghemOmic_Inhomog,PhysRevE.90.012810}, with different effective infection rate $\tau_m$ for each community $\mathcal{N}_m$. We use NIMFA and obtain for $\forall m =1,2,\ldots,M$,
\small{
\begin{align}
v_{\infty}^{(\mathcal{N}_m)} (n_m,u_{\infty})=1-\frac{1}{1+\tau _{m}(n_{m}-1) v_{\infty}^{(\mathcal{N}_m)} +\tau _{m}u_{\infty }}%(N_j,u_{\infty})
\label{steady_state_NIMFA_node_i_1}
\end{align}
}\normalsize where $v_{\infty}^{(\mathcal{N}_m)}$ is the metastable state infection probability for any non-core node of community $\mathcal{N}_m$ and $u_{\infty}$ is the infection probability for the core node. $v_{\infty}^{(\mathcal{N}_m)}$ is a function of $n_m$ and $u_{\infty}$.

All three considered classes of networks have a common property: with the exception of the \emph{core node} in the multi-communities network, if a node decides to invest in antivirus protection, the resulting (induced) overlay graph still belongs to the same class as the initial graph, only with less nodes. There are not many such classes of networks and this property is crucial in the game-theoretic analysis.

\section{Game model on a single community network}\label{sec:GameModelComplete}
In the investment game on the complete graph $K_{N}$, each node is a player and decides individually to invest in antivirus protection. The investment cost is $C$, while the infection cost is $H$. When a node invests, it is assumed to be directly immune to the virus and not part of the epidemic process anymore. Hence, this node cannot infect other nodes nor can be infected. If a node does not invest in antivirus protection, it is prone to the epidemics and might be infected by the virus (with rate $\beta$), but also can use additional protective mechanisms, like recovery or anti-spyware software (with rate $\delta$). The induced network, without the nodes that decide to invest, is also a complete graph and it influences on the epidemic spread process.
%The number of nodes that do not invest is $n$.

\subsection{Pure strategies}
The investment cost for any player is a constant $C$ and does not depend on the action of the other players. If a player $i$ decides not to invest, his cost is a linear function of its infection probability $v_{i,\infty}(n)$ in the metastable state of the SIS process. The probability $v_{i,\infty}(n)$ depends explicitly on the number of nodes $n$ that decide not to invest. In other words, there is an initial contact graph $G=K_N$ in which all the nodes are connected and the decisions of all the nodes induce an overlay graph $K_n$ only composed of the nodes that have decided \emph{not} to invest.

%old
%The investment cost for any player is a constant $C$ and does not depend on the action of the other players. If a player decides not to invest, his cost is a linear function of his expected infection time\small{
%\begin{align*}
%E[T] = \frac{1}{T}\int_{0}^{T}E[X_i(N;t)]dt = \frac{1}{T}\int_{0}^{T}\text{Pr}[X_i(N;t)=1]dt,
%\end{align*}
%}\normalsize since $E[X_i ] = \Pr [X_i]$ for Bernoulli random variables. 
%%Each node of the network (graph) is free to decide whether or not to invest in an antivirus protection. 
%
%When the time $T$ becomes large, $E[T]$ is approximated by the infection probability $v_{i,\infty}(n)$ of node $i$  in the metastable state of the SIS process. The probability $v_{i,\infty}(n)$ depends explicitly on the number of nodes $n$ that decide not to invest. In other words, there is an initial contact graph $G$ in which all the nodes are connected and the decisions of all the nodes induce an overlay graph $G_g = K_n$ only composed of the nodes that have decided \emph{not} to invest.

\subsubsection{Congestion Game}
%Due to players' decisions, we have a congestion game, because the cost (negative utility) of each player depends on the number of players that have decided not to invest. 
Each node has the choice between two actions: invest (further denoted by $1$) or not (further denoted by $0$). The negative utility of a player, in case he does not invest, depends on the number of players that choose the same action ($0$) not to invest. We denote by $\sigma_i \in \{0,1\}$ the action of node $i$. For example, the
cost $S_{i1}$ of a player $i \in \{1,2,\dots,N\}$ which decides to invest is defined by: $S_{i1}=C:=S_1$, while the cost of a player $i$ which decides not to invest is: $S_{i0}(n)=H v_{i,\infty} (n) :=S_0(n)$. This game is a congestion game~\cite{Rosenthal73} as the cost of a player depends on the number of players that choose his action. In the context of a congestion game, a (pure) Nash equilibrium is a vector of (pure) strategies, characterized by the number of nodes $n^*$ that do not invest. We remark that several Nash equilibria lead to the same $n^*$. The conditions for pure Nash Equilibrium are given in Definition~\ref{def:NashEquilibrium}. %states We are interested in the existence and uniqueness of this value $n^*$.
%, where $n$ is the number of players that do not invest

\begin{definition}\label{def:NashEquilibrium}
At a Nash equilibrium, no node has an interest to change unilaterally his decision. The number $n^*$  of nodes that do not invest at a Nash equilibrium is defined for any player $i$, by: $S_{i1} \leq S_{i0}(n^*+1)$ and $S_{i0}(n^*) \leq S_{i1}$.
%\small{
%\begin{align*}
%S_{i1} &\leq S_{i0}(n^*+1)\quad \mbox{and}\quad
%S_{i0}(n^*) \leq S_{i1}.
%\end{align*}
%}\normalsize
\end{definition}
Our game is symmetric as all players share the same set of cost functions. The following important property (in Proposition~\ref{prep:potential}) says that our game is not only a congestion game but also a potential game, due to the potential formula in~\cite[Theorem 3.1]{Monderer96}. As mentioned in Section~\ref{SIScompleteGraph}, we denote $v_{\infty} = v_{i,\infty}$ for $\forall i$ in this case.

\begin{proposition}\label{prep:potential}
The game is a \emph{potential game}, where $\Phi(n)=C (N-n)+H\sum_{j=2}^{n}v_{\infty} (j)$ is the potential function of the game.
%\small{
%\begin{align*}
%\Phi(n)=C (N-n)+H\sum_{k=2}^{n}v_{\infty} (k)
%\end{align*}
%}\normalsize is the potential function of the game.
\end{proposition}
%
%\begin{proof}
%The proof follows directly from the potential formula in Theorem 3.1 in~\cite{Monderer96} for any congestion game.
%\end{proof}

%\begin{proof}
%\ifthenelse{\boolean{Longversion}}{
%As our game is a congestion game, according to~\cite{Monderer96}, it is also a potential game. Moreover, Monderer and Shapley~\cite{Monderer96} define the potential function for this particular type of game. This function is the summation over all the different actions of the cost depending on all the possible repartition of the users. In our setting, by using the definition of the potential function given in \cite{Monderer96}, our potential function depends on the number $n$ of users that decide not to invest $
%\Phi(n)=C (N-n)+H\sum_{k=2}^{n}v_{\infty} (k)
%$.}
%{The proof could be found in our technical Report~\cite{}.}
%\end{proof}

The existence of a potential function in a game shows the existence of pure Nash equilibrium: any minimum of the potential function $\Phi$ is a pure equilibrium. The existence also allows decentralized procedures like best response dynamics or reinforcement learning~\cite{Chen04,Cominetti10} to converge to the pure Nash equilibrium. We can assume, for example, that an investment is valid only for a fixed amount of time and then each node pays again after expiration of his license. The Nash Equilibrium is fully determined by the number of nodes $n^*$ that do not invest and this is given in Proposition~\ref{prep:equilibriumBounds}.

\begin{proposition}\label{prep:equilibriumBounds}
For the number of nodes $n^*$ that do not invest at equilibrium, the following inequality holds:
\small{
\begin{align*}
v_{\infty} (n^*) &\leq \frac{C}{H}\leq v_{\infty} (n^*+1).
\end{align*}
}\normalsize Moreover, above the epidemic threshold ($\tau>\frac{1}{N-1}$), $n^*$ is uniquely defined by:
%$$
\small{
\begin{equation}
n^* =
\begin{array}\{{lc}.
\min \left \{ N,\lceil \frac{1}{(1-\frac{C}{H})\tau} \rceil \right \}, & \mbox{if }C<H\\ 
N,& \mbox{otherwise}
\label{criteriaEquilibrium}
\end{array}
\end{equation}
%n^*=\min \left \{ N,\lceil \frac{1}{(1-\frac{C}{H})\tau} \rceil \right \}.
%$$
}\normalsize where $\lceil x \rceil $ is the closest integer greater or equal than $x$ and $N$ is the total number of nodes.
\end{proposition}

\subsubsection{Performance of the equilibrium}

In order to evaluate the performance of the system, considering a non-cooperative behavior of each node, we use the Price of Anarchy (PoA) metric~\cite{NisanRoughgardenTardosVazirani(AlgoGameTheory)07}. We define the social cost $S(n)$ of this system, when $n$ users do not invest, as the summation of the cost for all users:
\small{
\begin{align}
S(n)=\sum_{i=1}^{N}S_{i\sigma_i}(n)=C(N-n)+nHv_{\infty} (n)
\label{SW_pure}
\end{align}
}\normalsize We define $n^{opt}$ such that:
$
n^{opt}=\arg\min_{n} S(n)
$, while the Price of Anarchy, considering pure strategies, is defined by: \small{
$$
PoA_p=\frac{S(n^*)}{S(n^{opt})}\geq 1.
$$
}\normalsize Before determining the Price of Anarchy, we characterize the globally optimal solution for the minimal social cost.
\begin{proposition}\label{prep:SocOptComGraph}
The value that minimizes the social cost is $n^{opt} \in \{N,\lceil 1+ \frac{1}{\tau}\rceil\}$.
\end{proposition}
%\begin{proof}
%\ifthenelse{\boolean{Longversion}}{
%The payoff of the nodes that do not invest is non-negative since $H \geq 0$ and $v_{\infty}(n) \ge 0$. If $n<1+\frac{1}{\tau}$,  $SW(n)=C(N-n)$ which decreases in $n$. On the other hand, if $n \geq 1+\frac{1}{\tau}$, the derivative of (\ref{SW_pure}) is $
%SW'(n)=H-C+\frac{H}{\tau (n-1)^2}
%$.
%
%Two cases can be distinguished:\\
%%\begin{itemize}
%1) $C<H$: the function $SW(n)$ is strictly increasing over the interval $[1+\frac{1}{\tau},N]$, so the minimum is achieved in $\lceil 1+ \frac{1}{\tau} \rceil$.\\
%2) $C \geq H$: the function $SW(n)$ is increasing over the interval $[1+\frac{1}{\tau},1+\sqrt{\frac{H}{\tau(C-H)}}]$ and decreasing over $[1+\sqrt{\frac{H}{\tau(C-H)}},N]$, so the minimum is achieved in $\{\lceil 1+ \frac{1}{\tau} \rceil,N\}$ depending on the parameters of the system.
%%\end{itemize}
%}
%{The proof could be found in our technical Report~\cite{}.}
%\end{proof}
Corollary~\ref{optimimEquilibrium} expresses the relation of the numbers of node that do not invest in a Nash Equilibrium and the optimal solution.
\begin{corollary}\label{optimimEquilibrium} The equilibrium value $n^*$ is at least as large as the optimum value $n^{opt}$, thus $n^* \ge n^{opt}$.
\end{corollary}
%\begin{proof}
%\ifthenelse{\boolean{Longversion}}{
%If  $C \geq H$, based on the proof in Proposition~\ref{prep:equilibriumBounds}, we have $n^{*} = N$, which is clearly as large as any value of $n^{opt} \le N$. Otherwise ($C<H$), based on Preposition~\ref{PoABound}: $ n^{opt} = \lceil 1+ \frac{1}{\tau} \rceil $ and $SW(n)$ is increasing. Using the definition of PoA: $SW(n^{opt})\leq SW(n^*)$, which gives $n^{opt} \le n^*$.
%}
%{The proof could be found in our technical Report~\cite{}.}
%\end{proof}
We have determined $n^{*}$ and $n^{opt}$ in Propositions~\ref{criteriaEquilibrium} and \ref{prep:SocOptComGraph}, respectively. Via (\ref{SW_pure}), we can find $PoA_p$ in an exact, but rather complex form. Moreover, we can obtain a simple upper bound for $PoA_p$.
\begin{corollary}\label{PoABoundComp}
The Price of Anarchy $PoA_p$ is bounded by:
\small{
\begin{align*}
 1 \le \text{$PoA_p$} \le \frac{1}{1-(1+\frac{1}{\tau})\frac{1}{N}}.
\end{align*}
}\normalsize
\end{corollary}
%\begin{proof}
%\ifthenelse{\boolean{Longversion}}{
%First, the denominator $SW(n^*)$ of $PoA_p$ is strictly lower than $CN$. Indeed, using Preposition~\ref{prep:equilibriumBounds} into (\ref{SW_pure}) we have: $SW(n^*)=C(N-n^*)+n^*Hv_{\infty} (n^*) = CN - n^* (C- Hv_{\infty} (n^*)) < CN$.
%
%In Proposition~\ref{prep:SocOpt}, we obtain $n^{opt} \in \{N,\lceil 1+ \frac{1}{\tau}\rceil\}$. If $n^{opt} = N$ then $n^* = N$ (Corollary~\ref{optimimEquilibrium}) and $PoA_p=1$. For the case $n^{opt} =\lceil 1+ \frac{1}{\tau}\rceil $, the following (based on Proposition~\ref{prep:SocOpt}) applies: (i) $C<H$: the function $SW(n)$ is strictly increasing and\\ (ii) $C \geq H$ and $n^{opt} =\lceil 1+ \frac{1}{\tau}\rceil$, $SW(n)$ is also strictly increasing. Therefore, in both cases $n^{opt} = \lceil 1+ \frac{1}{\tau} \rceil \ge 1+ \frac{1}{\tau}$, hence $SW(n^{opt}) \ge SW(1+ \frac{1}{\tau}) = C(N-(1+ \frac{1}{\tau}))$ i.e. $\text{$PoA_p$}=\frac{SW(n^{opt})}{SW(n^*)} \ge 1-(1+\frac{1}{\tau})\frac{1}{N}$.
%%\small{
%%\begin{align*}
%%\text{$PoA_p$}&=\frac{SW(n^{opt})}{SW(n^*)} \ge 1-(1+\frac{1}{\tau})\frac{1}{N}.
%%\end{align*}}
%}{The proof could be found in our technical Report~\cite{}.}\end{proof}

\subsubsection{Decentralized Algorithm}
%We consider discrete time slots and we denote by $p_i(t)$ the probability that node $i$ invests at time slot $t$. 

Here, we propose a simple fully decentralized Reinforcement Learning
Algorithm (RLA) that converges to a pure Nash Equilibrium in our invest game. At each discrete time slot $k$, independently, each node $i$ decides whether to invest in antivirus protection. We denote by $\bar{\sigma}[k]=(\sigma_1[k],\ldots,\sigma_N[k])$ the vector of pure actions of all the nodes at time $k$. The pure action $\sigma_i[k]$ of node $i$ at time slot $k$ is an element from $\{0,1\}$, where action $1$ means node $i$ invests and action $0$ otherwise. The probability that node $i$ invests at time slot $k$ (i.e. $\sigma_i[k]=1$) is denoted by $p_i[k]  = \Pr [\sigma_i[k]=1]$. The corresponding induced complete graph is $K_{n}$, where $n[k]=N- \sum_{j=1}^N \sigma_j[k]$ is the number of nodes that do not to invest. Our learning algorithm is the following:
%
%  with probability $p_i[k] $. We denote by $\bar{\sigma}[k]=(\sigma_1[k],\ldots,\sigma_N[k])$ the vector of pure actions of all the nodes at time $k$. The pure action $\sigma_i[k]$ of node $t$ at time slot $k$ is an element from $\{0,1\}$, where (i) action $1$ means node $i$ invests and (ii) action $0$ otherwise. The probability of $\sigma_i[k]=1$
%
%    (for any time $t$, $\sigma_i(t)=1$ if node $i$ invests at time $t$, it is 0 otherwise). More precisely, the probability $p_i (t)$ determines whether the action of player $i$ is $\sigma_i (t)=1$, for example (a) if $p_i (t)=1$ then $\sigma_i (t)=1$; (b) if $p_i (t)=0$ then $\sigma_i (t)=0$; (c) if $p_i (t)\in (0,1)$ then $\sigma_i (t)\in \{0,1\}$ based on $p_i (t)$. $G_g(\bar{\sigma}(t))$ is the corresponding induced graph composed only of the nodes that do not to invest. The principle of our learning algorithm is the following:
\begin{enumerate}
\item Set an initial probability $p_i[0]$, for each user $i\in \{1,\ldots,N\}$.
\item At every time slot $k$, each node $i$ invests with probability $p_i[k] = \Pr [\sigma_i[k]=1]$, which determines its pure action $\sigma_i[k]$.
\item Each player $i$ has a negative utility (cost) $S_{i\sigma_i[k]}(n[k])$, which is equal to:
\small{
\begin{align*}
S_{i\sigma_i[k]}(n[k]) =
\begin{array}\{{cl}.
C,& \hspace{-0.8em}\mbox{if} \quad \sigma_i[k]=1,\\
Hv_{i,\infty}(n[k]),&\hspace{-0.8em}\mbox{otherwise.}\\
\end{array}
\end{align*}
}\normalsize where $n[k]=N- \sum_{j=1}^N \sigma_j[k]$.
\item The cost of each node $i$ is normalized:
$
\tilde{S}_{i\sigma_i[k]}(n[k])=\frac{S_{i,\sigma_i[k]}(n[k])}{C+H}.
$
A normalization is necessary for this algorithm~\cite{Sastry94}.
\item Each node $i$ updates its probability according to the following rule:
\small{
\begin{align*}
p_i[k+1] \leftarrow p_i[k]+b[k]\tilde{S}_{i\sigma_i[k]}(n[k])(\sigma_i[k]-p_i[k]),
\end{align*}
}\normalsize where $b[k]$ is the learning rate.
\item Stop when a stopping criterion is met (for example, the maximum of the differences between consecutive updates is smaller than a small $\varepsilon$); else increase $k$ by 1 and go to step 2).
\end{enumerate}

Sastry \emph{et al.}~\cite{Sastry94} have proved that as $b[k]\rightarrow 0 $, for any node $i$, the discrete probability sequence $p_i[k]$ converges weakly to the solution $p_i (t)$ of the following continuous time ordinary differential equation (ODE)
\small{
\begin{align*}
\dot{p}_i (t)&= p_i (t)(1-p_i (t)) \big[S_{i0}\big(n(p_i (t))\big)-S_{i1}\big(n(p_i (t))\big)\big],
\end{align*}
}\normalsize which is known in evolutionary game theory as the \emph{replicator dynamic equation}~\cite{weibull}. Bournez and Cohen~\cite{Bournez2013} show that any potential game possesses a Lyapunov function $F$. Hence, our decentralized algorithm converges almost surely to a pure ($\varepsilon$-)Nash equilibrium of our game (see e.g.,~\cite{Kleinberg:2009:MUO:1536414.1536487,Bournez2013}). Due to~\cite[Theorem 5]{Bournez2013}, the convergence time $T \le O(\frac{F(\sigma[0])}{\varepsilon})$ only depends on $\varepsilon$ and the Lyapunov value $F(\sigma[0])$ of the initial strategy $\sigma[0]$. The algorithm is simple and fully distributed, the only required information for each node is its instantaneous cost at each time slot. Similar algorithms have been successfully used in many networking applications~\cite{Coucheney09,Xin2008} and we illustrate the behavior of our algorithm in Section~\ref{sec:Simulations}.
\vspace{-0.2cm}
\subsection{Symmetric mixed strategies}
We now assume that each individual decides with a probability $p$ to invest in the anti-virus protection. Moreover, the game is symmetric and then we look for a symmetric mixed Nash equilibrium. Each individual is faced with a new game, which depends on the realization of the random choice process of all the other individuals. We denote by $\bar{S}_{i\sigma_i}(p)$ the expected cost of player $i$ choosing the pure strategy $\sigma_i$ against the probability choice $p$ of the other $N-1$ players. For any user $i$, we have $
\bar{S}_{i\sigma_i}(p)=\sum_{n=0}^{N-1}S_{i\sigma_i}(n+1)\binom{N-1}{n}(1-p)^n p^{N-1-n}
$, where by definition $S_{i0}(1)=0$. Hence, the total expected cost of node $i$ which invests with probability $p'$ and when all the other nodes invest with probability $p$, is:
\small{
\begin{align}
\bar{S}_{i}(p',p)=p'\bar{S}_{i1}(p)+(1-p')\bar{S}_{i0}(p). \label{mixedDefinition}
\end{align}
}\normalsize

\begin{definition}\label{def:equilibrium}
(indifference property) At equilibrium, the probability $p^*$, that a node invests, is the solution of $
\bar{S}_{i}(0,p^*)=\bar{S}_{i}(1,p^*)
$.
\end{definition}
Definition~\ref{def:equilibrium} is a starting point for the characterization of the mixed equilibrium. The existence and uniqueness of a symmetric mixed equilibrium $p^*$ are shown by Propositions~\ref{prep:ExistUniqComp} and \ref{prep:MixedUniqEquil}, respectively.

\begin{proposition}\label{prep:ExistUniqComp}
A symmetric mixed equilibrium exists.
\end{proposition}

%\begin{proof}
%\ifthenelse{\boolean{Longversion}}{
%For any $p \in [0,1]$ and any player $i$, we have: $
%\bar{S}_{i}(1,p)=C\sum_{n=0}^{N-1}\binom{N-1}{n}(1-p)^np^{N-1-n}=C
%$. We also have: $
%\bar{S}_{i}(0,0)=H(1-\frac{1}{(N-1)\tau})>0
%$, and $\bar{S}_{i}(0,1)=0$.
%
%If $C<H(1-\frac{1}{(N-1)\tau})$ the mixed strategy $p^*$ exists because the function $\bar{S}_{i}(0,p)$ is continuous. Otherwise, we have for all $p \in [0,1]$, $\bar{S}_{i}(1,p)>\bar{S}_{i}(0,p)$, meaning that the strategy 0 is dominant irrespective of the mixed strategy of the other players. In this case, the action 0 is the equilibrium.
%}
%{The proof could be found in our technical Report~\cite{}.}
%\end{proof}
The equilibrium point $p^*$ can be determined from an exact, but rather complex, non-closed expression in $p$:
\small{
\begin{align}
\quad \bar{S}_{i}(0,p) =&\sum_{n=0}^{N-1}S_{i0}(n+1) \binom{N-1}{n} (1-p)^np^{N-1-n} \notag\\
%=&H\sum_{n=0}^{N-1}v_{\infty} (n+1) \binom{N-1}{n} (1-p)^np^{N-1-n}, \notag \\
=&H\sum_{n=\underline{n}}^{N-1}(1-\frac{1}{\tau n}) \binom{N-1}{n} (1-p)^np^{N-1-n}, \label{mixedTemp}
\end{align}
}\normalsize with $\underline{n}=\lceil \frac{1}{\tau}\rceil $ because $S_{i0}(n+1) = v_{i,\infty} (n+1)$ if $\tau \ge \frac{1}{n}$ and $S_{i0}(n+1) = 0$, otherwise. 

\begin{proposition}\label{prep:MixedUniqEquil}
The symmetric mixed equilibrium is unique.
%The equilibrium at $p^*$ is unique.
\end{proposition}
Expression (\ref{mixedTemp}) involves \emph{generalized hyper-geometric functions}~\cite{gasper2004basic}, which explains the difficulty of finding a closed form for $p^{*}$.

\subsubsection{Approximation}
In order to get a closed-form expression of the symmetric mixed strategy, we consider the following approximation: instead of considering a player faced to realize a symmetric mixed strategy of the other players and optimizing his average cost, we consider that a player is part of an average game. If player $i$ chooses strategy 1 with probability $p'$ we obtain the following average approximated cost: 
\small{
\begin{align}\label{mixedApprox}
\hat{S}_i^{\text{approx}}(p',p)=p'C+(1-p')Hv_{i,\infty}(\bar{n}(p)+1),
\end{align}
}\normalsize where $\bar{n}(p)$ is the average number of nodes, except node $i$, that decide not to invest, i.e. $\bar{n}(p)=(1-p)(N-1)$.
%$C_{n}^{N-1} (x) = \binom{N-1}{n} x^n (1-x)^{N-1-n}$, for $n=0,\dots,N-1$\emph{Bernstein Basis Polynomials},

We denote by $B_{N}(f;x) = \sum_{n=0}^{N} f(\frac{n}{N})\binom{N}{n} x^n (1-x)^{N-n}$ the $N$-th \emph{Bernstein Polynomial} on function $f$, see e.g.~\cite{lorentz1953bernstein}. The following Theorem~\ref{Bernstein1} is due to Bernstein~\cite{BernsteinApprox} and is crucial for proving that the approximation works.
\begin{theorem}\label{Bernstein1} 
(Bernstein, 1912) If f(x) is a continuous and bounded function defined on $x\in [0, 1]$, then for each $\varepsilon >0$ there is a positive integer $n_0 (\varepsilon)$ such that
\small{
\begin{align*}
\mid f(x) - B_{N-1}(f;x) \mid < \varepsilon
\end{align*}
}\normalsize for all $x\in [0, 1]$ and $N \ge n_0 (\varepsilon)$.
\end{theorem}
Applying Theorem~\ref{Bernstein1} for $f(x) = p'C+(1-p')H(1-\frac{1}{\tau (N-1) x})$ if $x \in [\frac{1}{\tau (N-1)},1]$ and $f(x)=p'C$ if  $x\in [0,\frac{1}{\tau (N-1)})$, using (\ref{mixedDefinition}), (\ref{mixedTemp}) and (\ref{mixedApprox}), taking $x = 1-p$, yield 
\small{
\begin{align*}
\mid \hat{S}_i^{\text{approx}}(p',p) - \bar{S}_{i}(p',p) \mid < \varepsilon.
\end{align*}
}\normalsize Therefore, for high enough $N \ge n_0 (\varepsilon)$, $\hat{S}_i^{\text{approx}}(p',p)$ can be arbitrary close, for any $\varepsilon,$ to the real $\bar{S}_{i}(p',p)$. A theoretical estimate of $n_0$ can be obtained, for example, following the proof of Theorem~\ref{Bernstein1} in~\cite{BernsteinApprox}. Moreover, numerical simulations in Fig.~\ref{poapure} show that the corresponding PoAs are similar even for low $N$. Finally, based on this approximation, we can characterize the mixed equilibrium in Proposition~\ref{prep:SymmetricMixedComp}.
%$\hat{S}_i^{\text{approx}}(p',p)$ and $\bar{S}_{i}(p',p)$ and

%    see e.g.~\cite{lorentz1953bernstein}

\begin{proposition}\label{prep:SymmetricMixedComp}
If we approximate the number of nodes that do not invest by its average, we obtain the following symmetric mixed Nash equilibrium is achieved for
\small{
\begin{align*}
\hat{p}^*=
\begin{array}\{{cl}.
1-\frac{H}{\tau (H-C)(N-1)},& \mbox{if} \quad C < H(1-\frac{1}{\tau (N-1)}),\\
0,&\mbox{otherwise.}\\
\end{array}
\end{align*}
}\normalsize and the corresponding social cost $S_{m}^{*} = N C$.
\end{proposition}
%
%\begin{proof}
%\ifthenelse{\boolean{Longversion}}{
%Using (\ref{infection_prob_node_i_regular_graph}), the average approximated cost is: $\hat{S}_i(p_i,p')= p_iC+(1-p_i)Hv_{i\infty}(\bar{n}+1) = p_iC+(1-p_i)H(1-\frac{1}{\tau (1-p') (N-1)})$.
%%\begin{align*}
%%\hat{S}_i(p_i,p')=&p_iC+(1-p_i)Hv_{i\infty}(\bar{n}+1)\\
%% %=& p_iC+(1-p_i)H(1-\frac{1}{\tau \bar{n}})\\
%% =& p_iC+(1-p_i)H(1-\frac{1}{\tau (1-p) (N-1)})
%%\end{align*}
%Based on Definition~\ref{def:equilibrium} for the equilibrium, assuming that is achieved for $p'=\hat{p}^{*}$, we have $\hat{S}_i(0,\hat{p}^{*})=\hat{S}_i(1,\hat{p}^{*})$, which gives $C= H(1-\frac{1}{\tau (1-p^{*}) (N -1)})$. Finally, $\hat{p}^{*} = 1-\frac{H}{\tau (H-C)(N-1)}$, if $C < H(1-\frac{1}{\tau (N-1)})$, otherwise $\hat{p}^{*} = 0$ is an equilibrium, according to Preposition~\ref{prep:ExistUniqComp}.
%%\begin{align*}
%%\hat{S}_i(0,\hat{p}^{*})=&\hat{S}_i(1,\hat{p}^{*}) \Leftrightarrow C=   H(1-\frac{1}{\tau (1-p^{*}) (N -1)}) \\
%%\hat{p}^{*} =& 1-\frac{H}{\tau (H-C)(N-1)}%1-\frac{H}{\tau (N-1)(H-C)}
%%\end{align*}
%}
%{The proof could be found in our technical Report~\cite{}.}
%\end{proof}

If the investment cost $C$ is higher than the curing cost $H$, then the equilibrium is $\hat{p}^*=0$, because even, if a node is infected, its cost $H$ is less than the cost $C$, then he would pay to be protected.

\subsubsection{Performance of symmetric mixed equilibrium}
The social cost can be defined considering the mixed strategies:\small{
\begin{align}
\hat{S}^{\text{approx}}(p) =\sum \limits_{i=1}^{N}\hat{S}_i^{\text{approx}}(p,p) \label{socCostApprox}%\bar{S}_i(p,p)
\end{align}
}\normalsize Further, using (\ref{mixedApprox}) into (\ref{socCostApprox}), we can compute the optimal social cost by finding the probability $\hat{p}^{opt}$:
\small{
\begin{align*}
&\hat{p}^{opt}=\arg\min \big[\hat{S}^{\text{approx}}(p)\big]\simeq\arg\min \big[\sum \limits_{i=1}^{N}\hat{S}_i^{\text{approx}}(p,p)\big]\\
&= \arg\min \big[ N \cdot (p C + (1-p)H\max \{0,1-\frac{1}{\tau (1-p) (N-1)}\}\big]\\
%=&\arg\min_p   \begin{array}\{{ll}.
%pC+(1-p)H(1-\frac{1}{\tau (1-p) (N-1)}),\mbox{ if} \quad p \in [0,1-\frac{1}{\tau(N-1)})&\\
%pC,\mbox{ if}\quad p \in [1-\frac{1}{\tau(N-1)},1]&
%\end{array}\\
&=\arg\min  N \cdot \begin{array}\{{ll}.
\hspace{-0.7em}(C-H)p+H(1-\frac{1}{\tau (N-1)}),\mbox{ if }  p \in [0,1-\frac{1}{\tau(N-1)})&\\
\hspace{-0.7em}pC,\mbox{ if } p \in [1-\frac{1}{\tau(N-1)},1].&
\end{array}
\end{align*}
}\normalsize and the solution is given in Proposition~\ref{prep:optimalAverage}.
\begin{proposition}\label{prep:optimalAverage} The optimal solution of the social cost is \\$N\cdot \min\{C,H\}(1-\frac{1}{\tau(N-1)})$
%\begin{align*}N\min\{C,H\}(1-\frac{1}{\tau(N-1)})
%\end{align*}
and it is achieved for
\small{
\begin{align*}
\hat{p}^{opt} = \begin{array}\{{ll}.
0,&\mbox{ if} \quad C>H\\
\text{$[$}0,1-\frac{1}{\tau(N-1)}\text{$]$},&\mbox{ if}\quad C=H\\
1-\frac{1}{\tau(N-1)},&\mbox{ if}\quad C<H.
\end{array}
\end{align*}
}\normalsize
\end{proposition}
%
%\begin{proof}
%\ifthenelse{\boolean{Longversion}}{
%First, the function is continuous in $p$, because the value is the same from the left and the right side of $1-\frac{1}{\tau(N-1)}$.  If $p\in[ 1-\frac{1}{\tau(N-1)},1]$ then it is increasing for any $C$ and $H$. If $C>H$, the function is increasing on the whole interval $p \in [0,1]$, hence $\hat{p}^{opt}=0$ and the value is $H(1-\frac{1}{\tau(N-1)})$. If $C=H$ the function is constant on $[0,1-\frac{1}{\tau(N-1)}]$, hence $\hat{p}^{opt} \in [0,1-\frac{1}{\tau(N-1)}]$ and the value is $H(1-\frac{1}{\tau(N-1)})$. If $C<H$ the function is decreasing on $[0,1-\frac{1}{\tau(N-1)}]$, hence $\hat{p}^{opt}=1-\frac{1}{\tau(N-1)}$ and the value is $C(1-\frac{1}{\tau(N-1)})$.
%}
%{The proof could be found in our technical Report~\cite{}.}
%\end{proof}
Based on Propositions~\ref{prep:SymmetricMixedComp} and \ref{prep:optimalAverage}, we approximate $PoA_m$ in the case of mixed strategies:
\begin{corollary}\label{PoAMixedCorr}
When each node uses a mixed strategy, the Price of Anarchy $PoA_m$ can be approximated by:
\small{
\begin{align*}
PoA_m \approx\frac{C}{\min\{C,H\}(1-\frac{1}{\tau(N-1)})}.
\end{align*}
}\normalsize
\end{corollary}
%\begin{proof}
%\ifthenelse{\boolean{Longversion}}{
%Based on Proposition~\ref{prep:optimalAverage} and $SW(p^*)=NC$.
%}
%{The proof could be found in our technical Report~\cite{}.}
%\end{proof}

\subsection{Comparison of strategies}
In the previous sections, we have studied two different approaches for our non-cooperative investment game: the pure and the mixed strategies. These two game variants assume significantly different decision processes for each node. First, the approximation of the expected number of nodes that do not invest at equilibrium is very close to the result obtained using the potential game approach: $
\hat{n}=N(1-\hat{p}^*)\simeq n^*$. Second, we compare the social costs obtained in each situation, and we observe that pure strategies always yields a lower social cost compared with symmetric mixed strategies.%We gave a simple decentralized algorithm that converges to the pure strategies equilibrium.

Based on Proposition~\ref{prep:SymmetricMixedComp}, $S_m^* =  CN$. On the other hand, in the proof of Corollary~\ref{PoABoundComp}, we find $S_p^* := S (n^*)>CN$. Corollary~\ref{qualityEquilibria} immediately follows.

\begin{corollary}\label{qualityEquilibria}
The social cost is smaller if all the nodes use a pure strategy ($S_p^*$) compared to the case in which all the nodes use a symmetric mixed strategy ($S_m^*$), i.e. $S_p^*<S_m^*$.
\end{corollary}

The bound achieved in Corollary~\ref{qualityEquilibria} is tight, because $(S (n^*)-CN)$ is small - based on Proposition~\ref{prep:equilibriumBounds}. This is also visualized in Fig.~\ref{poapure} in Section~\ref{sec:Simulations}, where indirectly, by comparing the Price of Anarchy for different equilibria, we show that the approximation leads to an almost correct value for the real expected cost.

\section{Game model in bipartite network}\label{sec:BipartiteNetwork}
In this section, we characterize the equilibrium points, their existence and uniqueness for the complete bipartite network $K_{M,N}$. If $m$ and $n$ nodes do not invest in an anti-virus, from partitions $\mathcal{M}$ and $\mathcal{N}$, respectively, the induced graph is also bipartite $K_{m,n}$. The results for the Nash Equilibria are given in Proposition~\ref{prep:equilibriumBoundsBipartite}.
\begin{proposition}\label{prep:equilibriumBoundsBipartite}
The equilibrium pair $(n^*,m^*)$ exists and satisfies the following inequalities. For each node
\small{
\begin{align*}
&\text{from }\mathcal{M}, \text{   } v_{\infty}^{(\mathcal{M})}(n^*,m^*)<\frac{C}{H}\leq v_{\infty}^{(\mathcal{M})}(n^*,m^*+1) \text{ and}\\
&\text{from } \mathcal{N},\text{   } v_{\infty}^{(\mathcal{N})}(n^*,m^*)<\frac{C}{H}\leq v_{\infty}^{(\mathcal{N})}(n^*+1,m^*)
\end{align*}
}\normalsize Moreover, above the epidemic threshold, the following hold:
\begin{enumerate}
\item for a given $n^*$ ($m^*$) there is no more than one $m^*$ ($n^*$).
\item for any $\tau$ and $\frac{C}{H}\ge\frac{1}{2}$; or $\tau \ge \frac{(H+C)(H-2C)}{2C(H-C)}$ and $\frac{C}{H}<\frac{1}{2}$: $|n^*-m^*| \le 1$ i.e. $n^*$ and $m^*$ are either equal or consecutive integers.
\item in general, it is possible to have multiple equilibria pairs such that $|n^*-m^*| \ge 2$ for some $(n^*,m^*)$.
\end{enumerate}
%
% holds $|n^*-m^*| \le 1$ or more precisely it is uniquely defined by:
%%$$
%\begin{equation*}
%(n^*,m^*) =\\%\{(v,v+1),(v+1,v)\}
%\begin{array}\{{ll}.
%(v+1,v+1),& \mbox{if $v$ is an integer.}\\
%(\lfloor v \rfloor,\lceil v \rceil)\text{ or }(\lceil v \rceil,\lfloor v \rfloor), & \mbox{otherwise.}
%\label{criteriaEquilibrium}
%\end{array}
%\end{equation*}
%%n^*=\min \left \{ N,\lceil \frac{1}{(1-\frac{C}{H})\tau} \rceil \right \}.
%%$$
%where $v = \frac{1}{(1-\frac{C}{H})\tau}$.
\end{proposition}
%\begin{proof}
%The proof is in Appendix~\ref{prep:equilibriumBoundsBipartiteProofApp}.
%\end{proof}
%When $n$ and $m$ players do not invest in anti-virus protection, from $\mathcal{N}$ and $\mathcal{M}$ partitions, respectively, the social welfare is given by:
The social cost is now given by:
\small{
\begin{align}
S(n,m)=&\sum_{i=1}^{N}S_{i\sigma_i}(n)=C(N-n+M-m)\nonumber \\
&+ H (n v_{\infty}^{(\mathcal{N})}(n,m) + m v_{\infty}^{(\mathcal{M})}(n,m)). 
\label{SW_pureBip}
\end{align}
}\normalsize We define the optimal pair $(n^{opt},m^{opt})$ as:
\small{
\begin{align*}
(n^{opt},m^{opt})=\arg\min_{(n,m)} S(n,m).
\end{align*}
}\normalsize and the Price of Anarchy: $PoA:=\frac{S(n^*,m^*)}{S(n^{opt},m^{opt})}$. Before proceeding with PoA, we first find the globally optimal solution in Proposition~\ref{prep:SocOptBipartite}.
%\small{
%\begin{align*}
%PoA:=\frac{SW(n^{opt},m^{opt})}{SW(n^*,m^*)}.
%\end{align*}
%}\normalsize

\begin{proposition}\label{prep:SocOptBipartite}
In $K_{N,M}$, the minimum (optimal) value of the social cost is equal to $S = \max\{\tau^2 MN-1,0\}\cdot\min \{\frac{C}{\tau^2 \max \{ M,N\}},H \frac{\tau(M+N)+2}{\tau(\tau M+1)(\tau N+1)}\}$. In particular,
\begin{enumerate}
\item if $M N \le \frac{1}{\tau^2}$, then $S=0$ and $(n^{opt},m^{opt}) = (N,M)$.
\item if $M N > \frac{1}{\tau^2}$, $ \tau \max \{ M,N\} \frac{\tau(M+N)+2}{(\tau M+1)(\tau N+1)} \ge \frac{C}{H} $ then: $S = C\frac{\tau^2 MN-1}{\tau^2 \max \{ M,N\}}$ and  $(n^{opt},m^{opt}) = (\frac{1}{\tau^2 M},M)$ if $M>N$; $(n^{opt},m^{opt}) = (N,\frac{1}{\tau^2 N})$ if $M<N$ or both points for $M=N$.
\item if $M N > \frac{1}{\tau^2}$, $ \tau \max \{ M,N\} \frac{\tau(M+N)+2}{(\tau M+1)(\tau N+1)} < \frac{C}{H} $ then $S = H \frac{(\tau^2MN-1)[\tau(M+N)+2]}{\tau(\tau M+1)(\tau N+1)}$ and $(n^{opt},m^{opt}) = (N,M)$.
\end{enumerate}
%\small{
%\begin{align*}
% &\{(1,\min\{\lfloor\frac {1}{\tau^2} \rfloor,M\}),(\min\{\lfloor\frac {1}{\tau^2} \rfloor,N\},1),(n',M),(N,m'),(N,M)\}
%\end{align*}
%} \normalsize depends on the parameters of the system, for which $SW$ is minimum, where $n' = $
\end{proposition}
Based on the results in Propositions~\ref{prep:equilibriumBoundsBipartite} and \ref{prep:SocOptBipartite}, we find a tight bound for the Price of Anarchy (PoA) in Corollary~\ref{PoABound}.
\begin{corollary}\label{PoABound}
The Price of Anarchy is bounded by:
\small{
\begin{align*}
&\text{PoA} \le \frac{\tau(M+N)}{\max\{\tau^2 MN-1,0\} \min \{\frac{1}{\tau \max \{ M,N\}},\frac{H(\tau(M+N)+2)}{C(\tau M+1)(\tau N+1)}\}}. %\\
%&\frac{\min\{C(M+N-1-\min\{N,M,\lfloor\frac{1}{\tau^2}\rfloor\}),\frac{H(MN\tau^2-1)}{\tau}(\frac{1}{\tau N+1}+\frac{1}{\tau M+1})\}}{C(N+M)}
\end{align*}
}\normalsize% where $k = \frac{C}{H}$.
\end{corollary}

The only used inequality in the proof of Corollary~\ref{PoABound} is from Proposition~\ref{prep:equilibriumBoundsBipartite}. Corollary~\ref{PoALowerBound} gives a better intuition for the bound in Corollary~\ref{PoABound}. %The lower bound in Corollary~\ref{PoABound} is never greater than $\min \{\frac{1}{2},\frac{H}{C}\}$.
\begin{corollary}\label{PoALowerBound}The upper bound in Corollary~\ref{PoABound} $\frac{\tau(M+N)}{\max\{\tau^2 MN-1,0\} \min \{\frac{1}{\tau \max \{ M,N\}},\frac{H(\tau(M+N)+2)}{C(\tau M+1)(\tau N+1)}\}}$ is greater than $\max \{2,\frac{C}{H} \} $.
\end{corollary}

When the bound of PoA from Corollary~\ref{PoABound} is accurate (i.e. close to the real PoA), Corollary~\ref{PoALowerBound} tells us that the social cost due to a decentralized investment decision is often twice larger than the optimal. In the bipartite network case, we talk about the number of nodes that do not invest in both partitions separately, which is more complex than the case of a complete graph (Section~\ref{sec:GameModelComplete}). Moreover, as will be shown later (Fig.~\ref{comparisonCompleteBipartite}), PoA is always smaller if the partitions are ``more balanced'' (same number of nodes) and for any bipartite graph with $N$ nodes, the PoA is always higher than the PoA of the complete graph with $N$ nodes. For a bipartite graph, not much can be said about the mixed equilibrium due the fact that the bipartite network is not symmetric, and a players' uniform social cost function cannot be defined.

\section{Game model in multi-communities network}\label{sec:CommunityNetwork}

In the single community case, we first show that pure equilibrium strategies yield better performance compared to symmetric mixed strategies. Hence, we decide to restrict the analysis of the multi-communities network investment game to the existence of a pure Nash equilibrium.% We start with the concept of \emph{parametric potential games}. 

%\subsection{Parametric potential games}\label{subsec:ParamPotComGames}

%Similarly as for the previous topologies, if $n_m$ nodes from the community $\mathcal{N}_m$ do not invest in an anti-virus, the induced graph $G_g$ is also a multi-communities network. 
In this section, we propose a new game theoretic concept, namely a \emph{parametric potential game}, which is defined as follows. We assume that the infection probability $u_\infty$ of the core node is given. Further, our game with $M$ communities is equivalent to $M$ independent potential games.
We observe that, if $u_{\infty}$ is given, then based on (\ref{steady_state_NIMFA_node_i_1}), the infection probability $v_{\infty}^{(\mathcal{N}_m)}$ of a non-core node in community $\mathcal{N}_m$
depends only on $n_m$, 
\small{
\begin{align}
v_{\infty}^{(\mathcal{N}_m)} (n_m,u_{\infty}) = \frac{V (\tau_{m},n_m, u_{\infty})\left( 1+\sqrt{1+\frac{4\tau
_{m}^{2}u_{\infty }(n_{m}-1)}{V (\tau_{m},n_m, u_{\infty})}}\right)}{2\tau _{m}(n_{m}-1)}\label{prep:ParamPotCommGame}
\end{align}%
}\normalsize where $V (\tau_{m},n_m, u_{\infty}) = \tau _{m}(n_{m}-1)-\tau _{m}u_{\infty}-1 $. Clearly, the resulting $u_{\infty } =1-\frac{1}{1+\sum_{m=1}^{M}\tau _{m}n_{m}v_{\infty}^{(\mathcal{N}_m)}} = 1 - \frac{1}{1+\sum_{m=1}^{M}n_m\frac{V (\tau_{m},n_m, u_{\infty}) +\sqrt{V (\tau_{m},n_m, u_{\infty})^{2}+4\tau _{m}^{2}u_{\infty }(n_{m}-1)}}{2(n_{m}-1)}}$ is in $(0,1)$, because the second term in the difference is positive and smaller than $1$. Hence, $u_{\infty }$ is feasible. The above mentioned expressions for $v_{\infty}^{(\mathcal{N}_m)} (n_m,u_{\infty})$ and $u_{\infty }$ are not related to the game and the derivations are given in Appendix~\ref{ProofsMultiComm}.

Further, nodes from each community play a potential game within their own community. The approach is the following: (1) we first compute the equilibrium $n_m^*$ in each community $\mathcal{N}_m$ using
the parametric potential game approach ($u_{\infty}$ fixed); (2) we then
compute the value of $v_{\infty}^{(\mathcal{N}_m)}$ by using the expression (\ref{prep:ParamPotCommGame}).

%\small{
%\begin{align*}
%\frac{1}{1+\sum_{m=1}^{M}n_m \frac{V (\tau_{m},n_m, u_{\infty}) +\sqrt{V (\tau_{m},n_m, u_{\infty})^{2}+4\tau _{m}^{2}u_{\infty }(n_{m}-1)}}{2(n_{m}-1)}}
%\end{align*}
%}\normalsize is positive and smaller than $1$. Hence, $u_{\infty }$ is feasible.
%, thus satisfies $u_{\infty }\in \left( 0,1\right) $. This is ensured by Proposition~\ref{prep:Feasibility}.
%%\small{
%%\begin{align}
%%&v_{\infty } =1-\frac{1}{1+\sum_{m=1}^{M}\tau _{m}N_{m}v_{m}} = 1 -  \notag \\
%%&\frac{1}{1+\sum_{m=1}^{M}\frac{N_m (V (\tau_{m},N_m, v_{\infty}) +\sqrt{V (\tau_{m},N_m, v_{\infty})^{2}+4\tau _{m}^{2}v_{\infty }(N_{m}-1)})}{2(N_{m}-1)}}
%%\label{Eq:feasible}
%%\end{align}%
%%}
%%\normalsize is satisfied for some $v_{\infty }\in \left( 0,1\right) $.
%Moreover, Proposition~\ref{prep:Feasibility} is important for the feasibility of a pure Nash equilibrium in this game.
%\begin{proposition}\label{prep:Feasibility}
%The parametric solution given in Preposition~\ref{prep:ParamPotCommGame} is feasible i.e. there exists a unique $u_{\infty} \in \left( 0,1\right)$.
%\end{proposition}
%We now describe an iterative procedure which enables us to find such pure Nash equilibrium.
\begin{figure*}[h!tb]
\centering

\subfloat[]{
\includegraphics[trim = 14mm 65mm 14mm 70mm,clip,width=0.3043\textwidth]{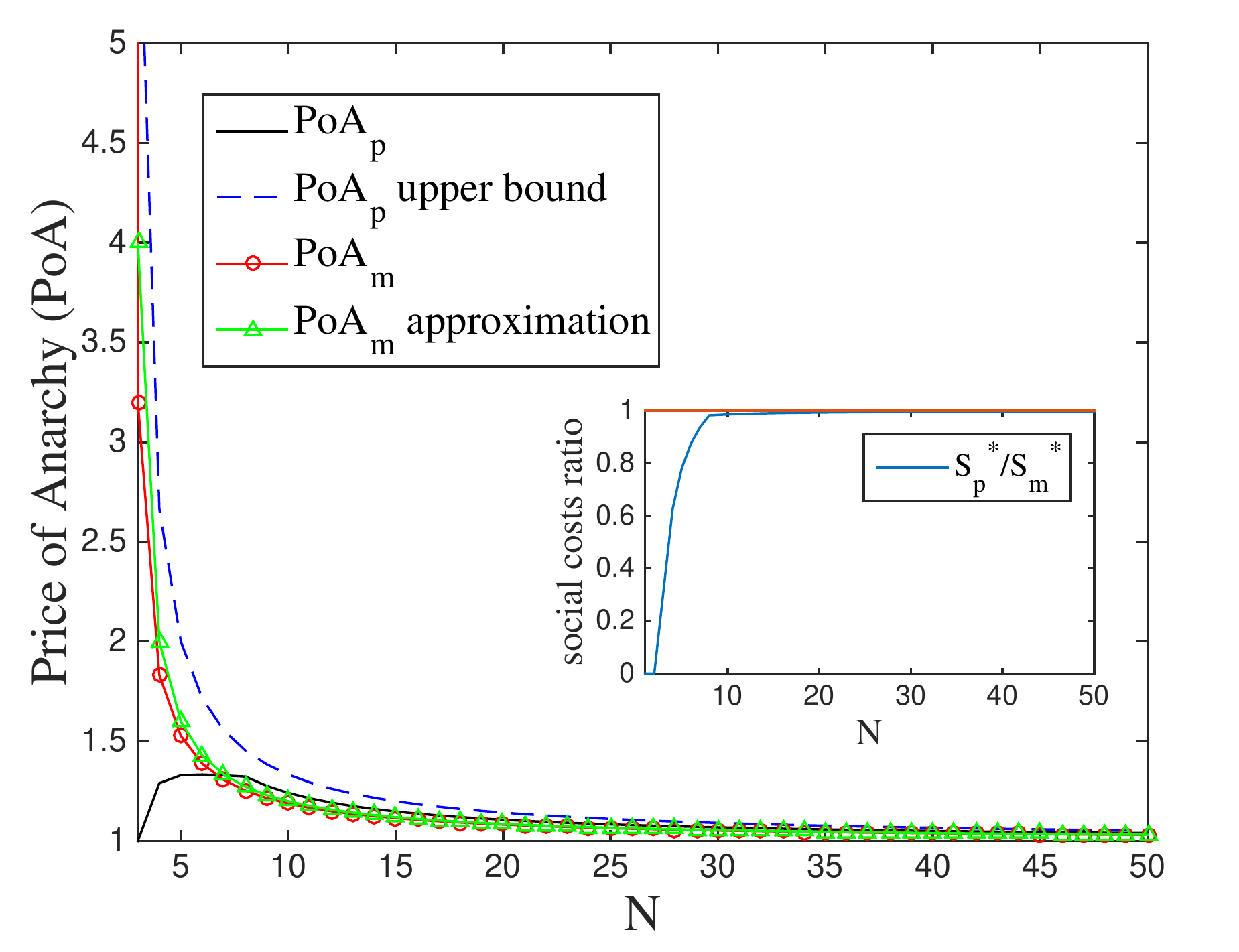}
\label{poapure}}
\subfloat[]{
\includegraphics[trim = 14mm 65mm 14mm 70mm,clip,width=0.3043\textwidth]{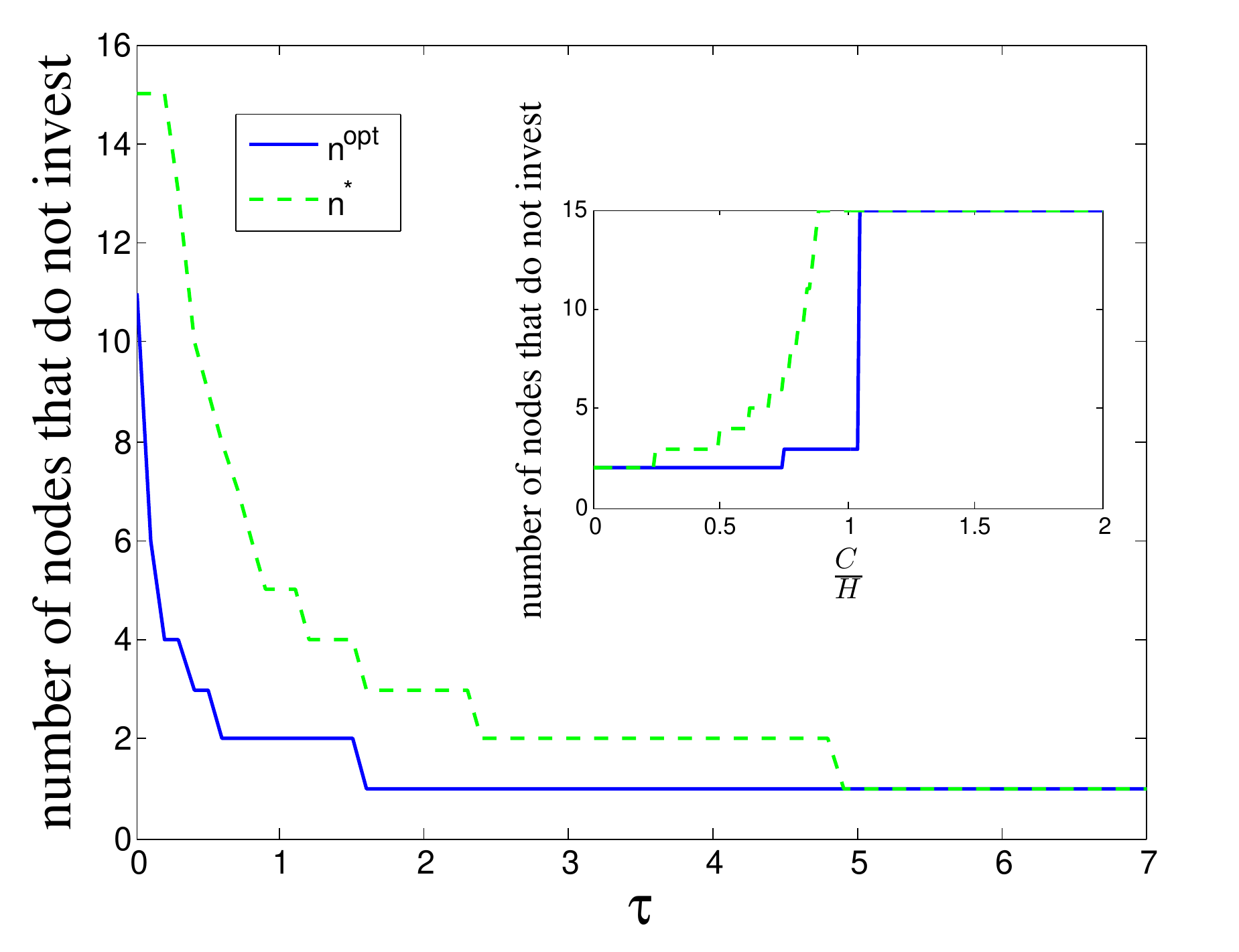}
\label{ntau}}
\subfloat[]{
\includegraphics[trim = 32mm 95mm 40mm 100mm,clip,width=0.3043\textwidth]{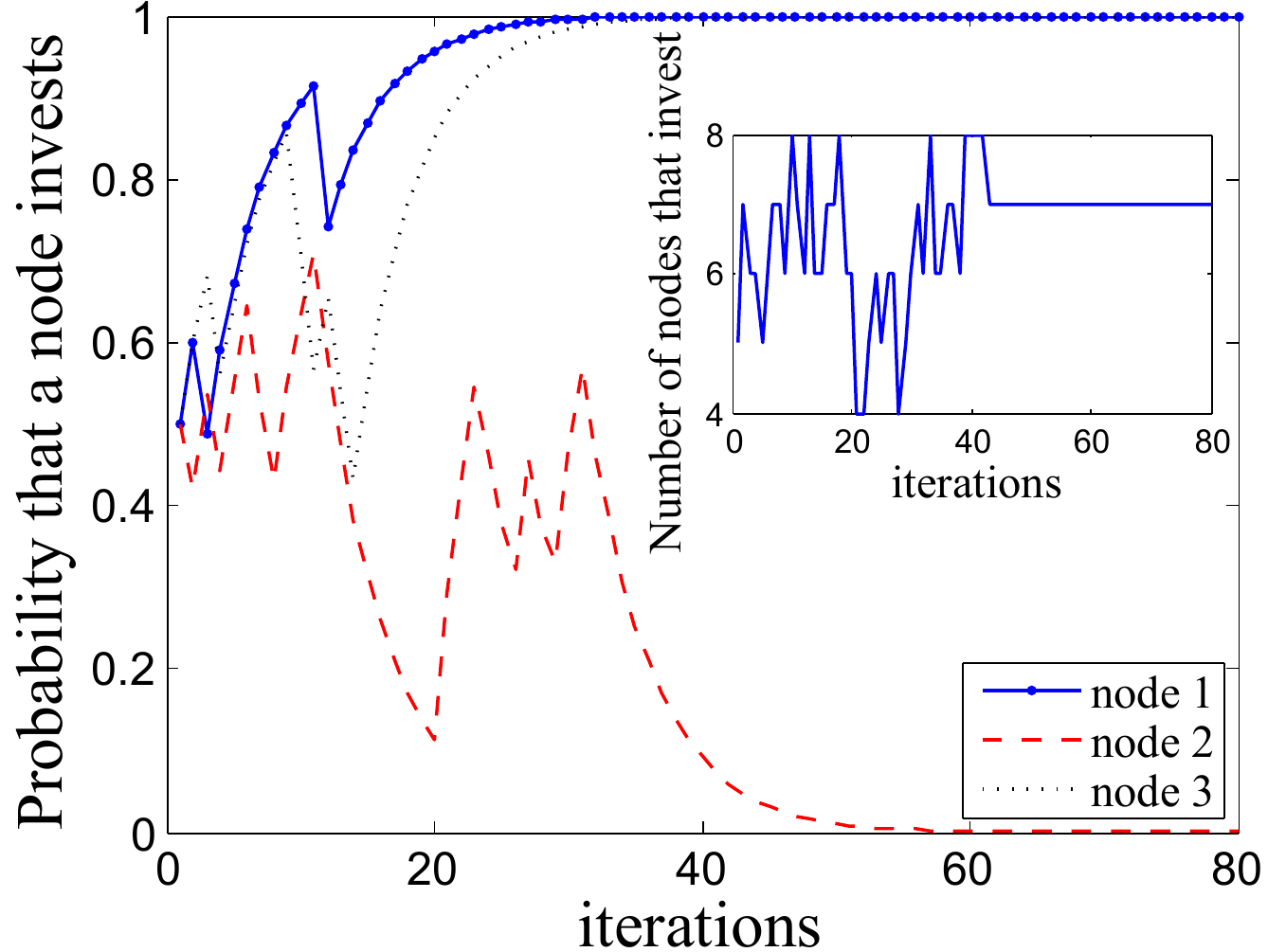}
\label{convRLA}}

\caption[]{(a) Price of Anarchy depending on the number of nodes $N$ (main plot). Ratio $\frac{S_p^*}{S_m^*}$ of the social costs depending on the size $N$ of the network (inset); (b) Number of nodes which do not invest as a function of the effective spreading factor $\tau$ (main plot) or the ratio $\frac{C}{H}$ (inset) for $N=15$; and (c) Convergence of the RLA to the pure Nash equilibrium. Probability of $3$ random nodes that invest at each iteration (main plot); number of nodes that invest after each iteration (inset).} \label{fig:regRunnTime}
\end{figure*}

%\subsection{Computation of the equilibrium}

We propose an iterative heuristic procedure to compute a pure Nash equilibrium of this parametric potential game.
\begin{enumerate}
\item Fixed an initial value for $u_{\infty}[0]$.
\item Based on this value, we solve the $M$ independent potential games and we obtain the solution vector\footnote{The iteration step $k$, given in brackets $[]$, is a discrete time and should not be mixed with the continuous time $t$ from Eq. (\ref{governEq:SIS}).} $\textbf{n}^*(u_{\infty}[k])=(n_{1}^*(u_{\infty}[k]),\ldots,n_{M}^*(u_{\infty}[k]))$, where $n_{m}^*(u_{\infty} [k])$ is the number of nodes of community $\mathcal{N}_m$ that do not invest at equilibrium in the $k$-th iteration, given the infection probability $u_{\infty}[k]$ of the core node in the $k$-th iteration. We denote for each community $\mathcal{N}_m$, the following parametric potential function: $\Phi_{m}(n_m,N_m,u_{\infty}[k])=C(N_m-n_m)+H\sum_{i=2}^{n_m}v_{\infty}^{(\mathcal{N}_m)}(i,u_{\infty}[k])$. Hence, $n_m^*(u_{\infty}[k]) =\arg\min_{n_m}\Phi_{m}(n_m,N_m,u_{\infty}[k])
$ for all $m$.
\item Further, we compute the infection probability of a node from community $m$ by the function $v_{\infty}^{(\mathcal{N}_m)}(n_m^*(u_{\infty}[k]),u_{\infty}[k])[k]$ from equation (\ref{prep:ParamPotCommGame}) and the infection probability of the core node: $
u_{\infty}[k+1]=1-\frac{1}{1+\sum_{m=1}^{M}\tau _{m}n_{m}^*(u_{\infty}[k]) v_{\infty}^{(\mathcal{N}_m)}(n_m^*(u_{\infty}[k]),u_{\infty}[k])[k]}$.
\item Stop if $\mid u_{\infty}[k+1]-u_{\infty}[k]\mid <\varepsilon$ and thus $u_{\infty}=u_{\infty}[k+1]$, otherwise increase $k \leftarrow k+1$ and start with step 2).
\end{enumerate}
The algorithm is a heuristic, a theoretical guarantee is not given, and it converges in practice as shown in Fig.~\ref{iter1anditerCN}. In Appendix~\ref{ProofsMultiComm}, we show that $u_{\infty}[k+1]$ is bounded from above and bellow by decreasing functions in $u_{\infty}[k]$, which influences on the convergence. In particular, if $u_{\infty}[k]$ is monotone sequence, then the convergence is proved.% as well as in the case if we neglect the ``integer rounding'' of $n_m^*(u_{\infty}^t)$.

%the convergence of all $v_{.,\infty}^{(\mathcal{N}_m)}$ ensure the convergence of $u_{\infty}$ and all $n_{m}^*$.

\section{Numerical evaluation}\label{sec:Simulations}

\subsection{Single-community network}

\subsubsection{Performance of the decentralized system}

We evaluate the performance of the decentralized system (equilibrium) compared to the centralized point of view (social optimum) via the Price of Anarchy of our system in different cases: pure and mixed strategies. We show how this metric depends on the system parameters, such as the number of nodes (decision makers), the effective epidemic spreading rate $\tau=\frac {\beta}{\delta}$ and the costs $C$ and $H$.
%\begin{figure}[h!tb]
%\begin{center}
%\includegraphics[trim = 32mm 95mm 40mm 100mm, clip,width=0.37\textwidth]{images/SWs_completeGraph.pdf}
%\end{center}
%\caption{Price of Anarchy depending on the number of nodes $N$ (main plot). Ratio $\frac{SW_p^*}{SW_m^*}$ of the social welfares depending on the size $N$ of the network (inset).}
%\label{poapure}
%\end{figure}

Fig.~\ref{poapure} illustrates the PoA with the following costs $C=0.4$, $H=0.5$ and the effective spreading rate $\tau=2/3$. We observe that when the number of nodes is relatively small ($N<8$): using pure strategies yields a smaller PoA compared to the case of mixed strategies. Moreover, we find that the upper bound of the pure $PoA_p$ is very close to both $PoA_p$ and $PoA_m$, when $N$ becomes relatively large ($N>10$). We also observe that the approximation of $PoA_{m}$, which is based on Corollary~\ref{PoAMixedCorr}, is very close to the exact $PoA_{m}$. In Fig.~\ref{poapure} (inset), we show that the ratio $\frac{S_p^*}{S_m^*}$ depends on the size $N$ of the network. Fig.~\ref{poapure} matches Corollary~\ref{qualityEquilibria}, i.e., the social cost obtained using pure strategies in the game, is lower than the one obtained via symmetric mixed strategies. This difference is noticeable when the network is small but diminishes quickly (e.g., for $N=8$, $\frac{S_p^*}{S_m^*}=0.9821$, and for $N\ge 10$, $\frac{S_p^*}{S_m^*}\approx 1$). 
%Finally, we have an important illustration saying that the PoA considering pure strategies is lower bounded, which is not the case considering symmetric mixed strategies. Particularly, in our case, the PoA is lower bounded by $0.75$ meaning that at worst considering a decentralized decision process induces $25\%$ loss of the individual payoff.

%\begin{figure}[tbh]
%\begin{center}
%\includegraphics[trim = 15mm 65mm 10mm 85mm, clip,width=0.37\textwidth]{images/ntau.pdf}
%\end{center}
%\caption{Number of nodes which do not invest as a function of the effective spreading factor $\tau$ (main plot) or the ration $\frac{C}{H}$ (inset plot) for $N=15$.}
%\label{ntau}
%\end{figure}

In Fig.~\ref{ntau}, we describe the number of nodes which do not invest considering the two methods: decentralized $n^{*}$ (Nash equilibrium) and the centralized case $n^{opt}$ (social cost), depending on the effective spreading rate $\tau$ (main plot) and ratio of the costs of investing and not investing $\frac{C}{H}$ (inset). First, we observe that our result is correct, i.e., considering a decentralized point of view, the number of nodes which invest is lower than that of the centralized point of view. This result is somewhat surprising, as in general in a decentralized system, the players are more suspicious and we would think that in our setting, more nodes would invest at equilibrium compared to the central decision. Second, those numbers are exponentially decreasing with the effective spreading rate $\tau$: the more the infection rate $\beta$ dominates the curing rate $\delta$, more nodes decide to invest in equilibrium. On the other hand, the number of nodes increases if the relative cost of investment decreases, as expected. However, the increase is faster in a decentralized system for a fixed $\frac{C}{H}$ (Fig.~\ref{ntau} inset).

\subsubsection{Distributed Learning}
We simulate the epidemic process in a full-mesh network of size $N=15$ in which each node $i$ uses the decentralized algorithm proposed in Section~\ref{sec:GameModelComplete}. The main problem is the two time scales: the update of strategies of each individual and the epidemic process. Our analysis is based on the metastable state of the infection process. When all nodes have taken their decision, the update rule of the RLA strategy depends on the infection probability for each node in the metastable state, i.e. $v_{\infty}$. We assume that the strategy update time scale is large compared to the infection time scale, and then, the infection probability has enough time to converge to the metastable state. This assumption is realistic in a case in which the decision to buy an anti-virus is not very often taken compared to the propagation process of virus in a network. % (\yh{speed of epidemics to metastable state})
%\begin{figure}[h!]
%\begin{center}
%\includegraphics[trim = 32mm 95mm 40mm 100mm, clip,width=0.37\textwidth]{images/RLAconvExtended.pdf}
%\end{center}
%\caption{Convergence of the RLA to the pure Nash equilibrium. Probability of $3$ random nodes that invest at each iteration (main plot); number of nodes that invest after each iteration (inset).}
%\label{convRLA}
%\end{figure}
In Fig.~\ref{convRLA} (main plot), we show the convergence of the RLA for three particular nodes. We observe that two nodes converge to the decision to invest and one not to invest. Moreover, in Fig.~\ref{convRLA} (inset) we plot the number of nodes that invest, at each iteration of our RLA. This number converges after few iterations to a pure equilibrium ($n^*=8$).
\begin{figure}[h!tb]
\centering

\subfloat[$\frac{C}{H} = 5$]{
\includegraphics[trim = 15mm 70mm 25mm 70mm,clip,width=0.237\textwidth]{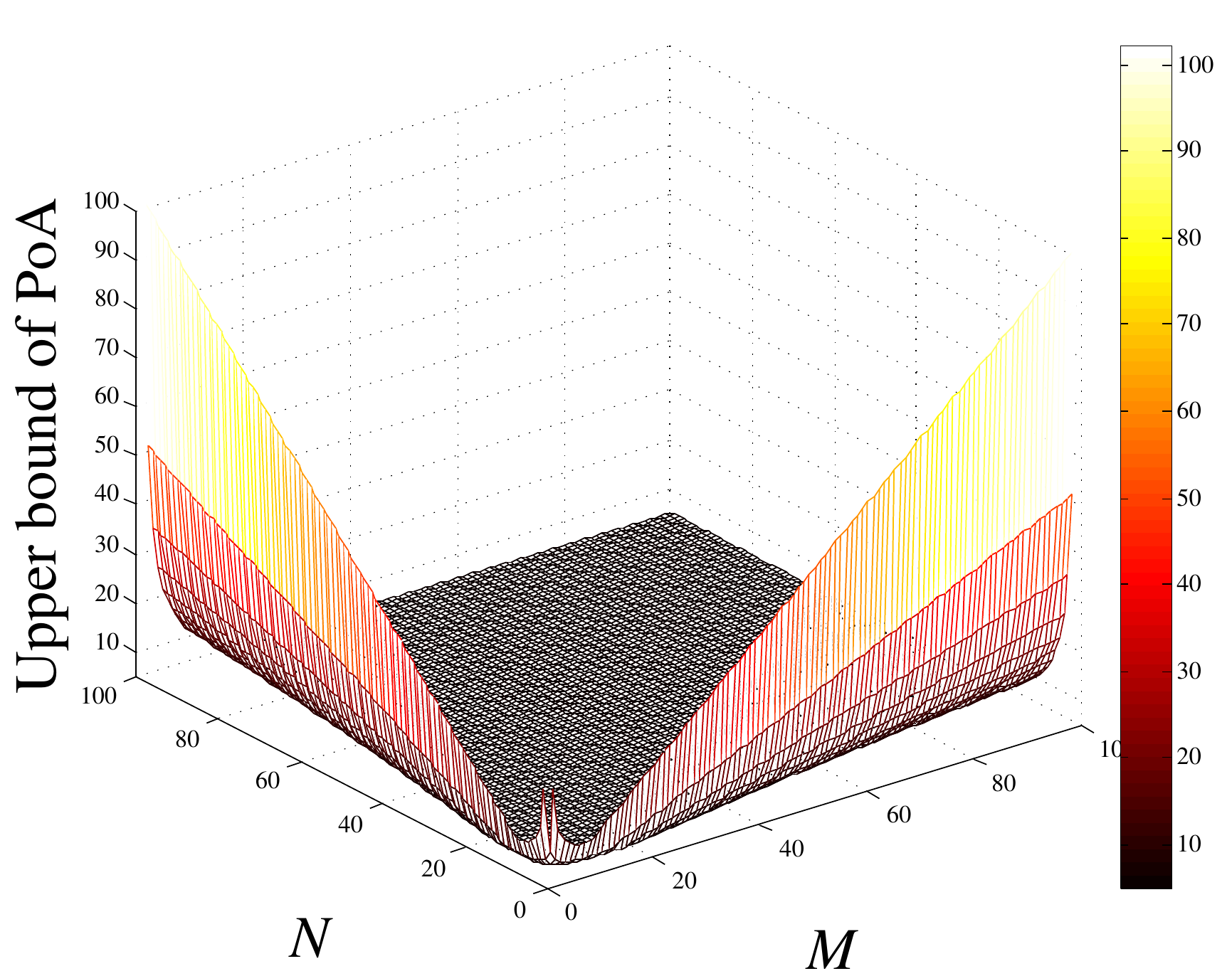}
\label{bipartite_3D5}}
\subfloat[$\frac{C}{H} = 5$]{
\includegraphics[trim = 12.8mm 68mm 24.8mm 68mm,clip,width=0.237\textwidth]{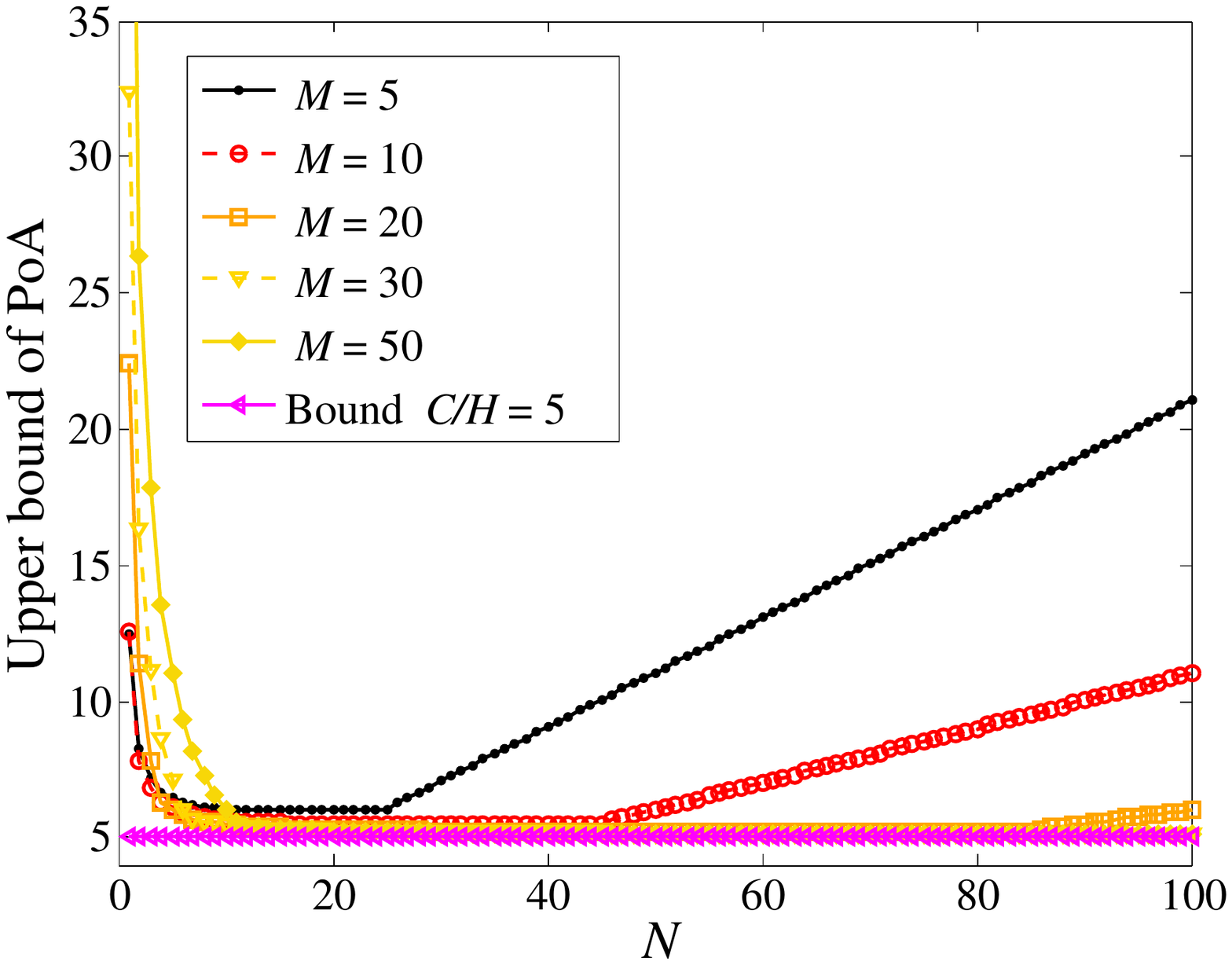}
\label{bipartite_2D5}}

\subfloat[$\frac{C}{H} = 0.2$]{
\includegraphics[trim = 15mm 70mm 25mm 70mm,clip,width=0.237\textwidth]{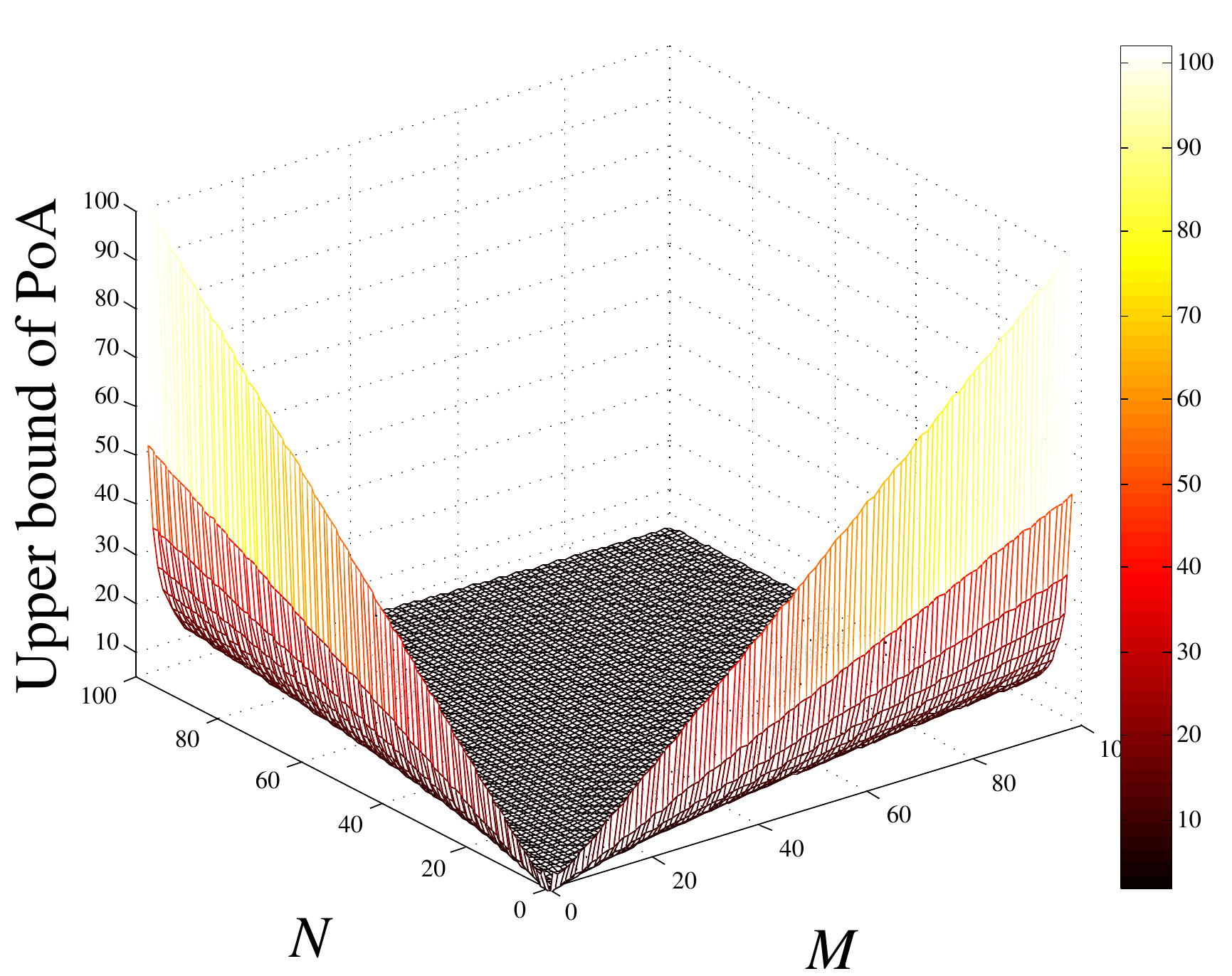}
\label{bipartite_3D02}}
\subfloat[$\frac{C}{H} = 0.2$]{
\includegraphics[trim = 12.8mm 68mm 24.8mm 68mm,clip,width=0.237\textwidth]{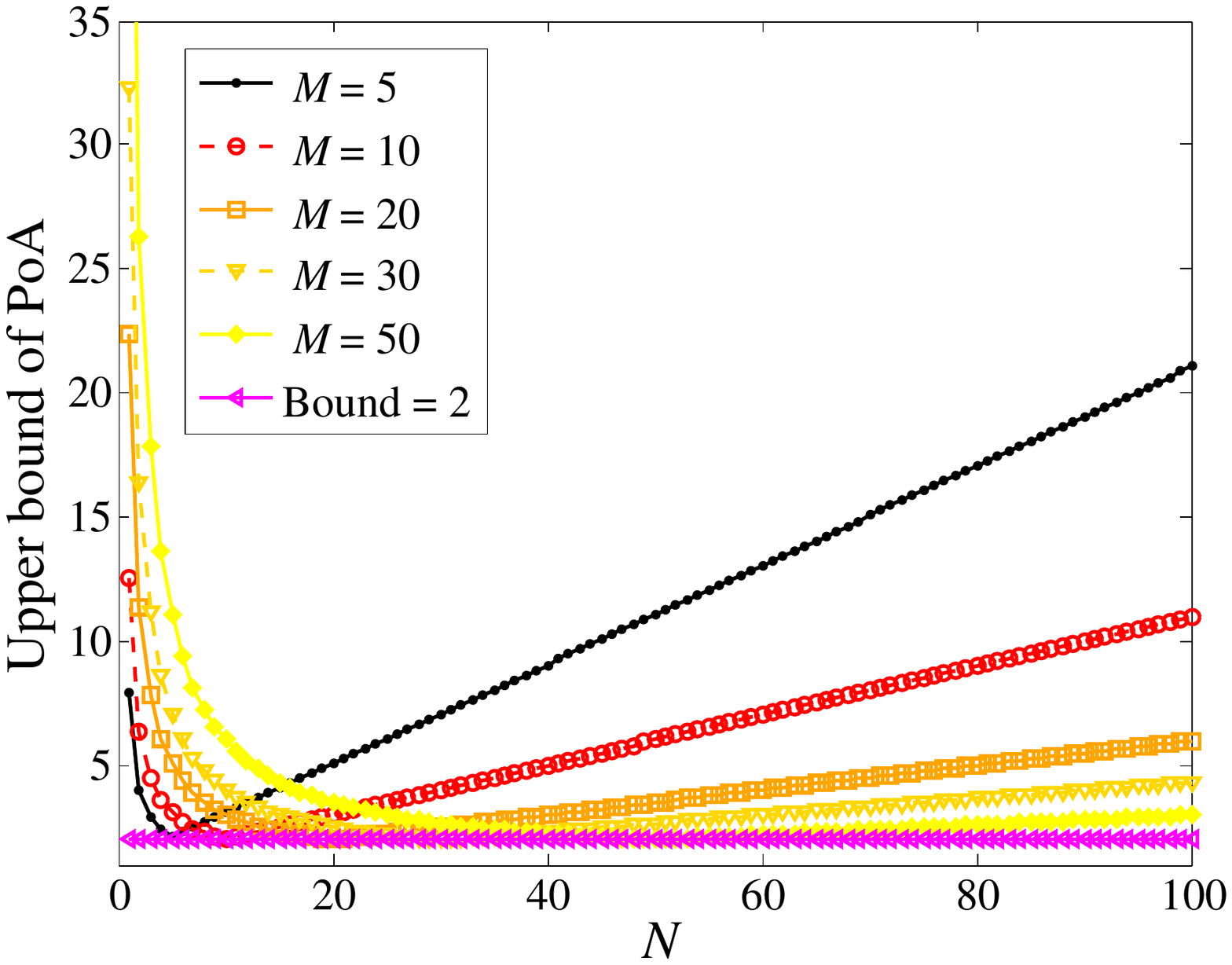}
\label{bipartite_2D02}}
\caption[]{The upper bound of the Price of Anarchy. (a) and (c) 3D plots as functions of $M$ and $N$. (b) and (d) 2D plots as functions of $N$ for fixed $M$.} \label{fig:bipartitePoA}
\end{figure}

\subsection{Bipartite network}
For the bipartite network, the upper bound of the Price of Anarchy (PoA) is illustrated in Fig.~\ref{fig:bipartitePoA}. In particular, Figs.~\ref{bipartite_3D5} and \ref{bipartite_3D02} show the upper bound as a function of both $M$ and $N$ as a $3$ dimensional plot, while Fig.~\ref{bipartite_2D5} and \ref{bipartite_2D02} demonstrate the change of the upper bound of the Price of Anarchy as a function of $N$ for several fixed values of $M$. All the figures confirm Corollary~\ref{PoALowerBound} that the upper bound of PoA is greater than the maximum of $2$ or $\frac{C}{H}$. In all the cases, the closer $M$ and $N$ are to one another - the smaller upper bound of PoA (black/dark regions in Figs.~\ref{bipartite_3D5} and \ref{bipartite_3D02} and the minimum values for the upper bound in Figs.~\ref{bipartite_2D5} and \ref{bipartite_2D02}). For fixed $M$ and $\frac{C}{H}<2$, the upper bound is dominated by $\frac{\tau^2 (M+N) \max\{M,N\}}{\tau^2 M N-1}$ (Corollary~\ref{PoABound}), which is a function that decreases in $N$ for $N<M$, achieves its minimum (close to $2$) and then increases for $N>M$ (Fig.~\ref{bipartite_2D02}). For fixed $M$ and $\frac{C}{H}>2$, the upper bound is dominated by $\frac{C}{H}\frac{\tau (M+N)(\tau M +1) (\tau N+1)}{(\tau^2 M N-1)(\tau(M+N)+2)}$ (Corollary~\ref{PoABound}), which is also a function that decreases in $N$ for $N<M$, achieves its minimum (close $\frac{C}{H}$) and stays almost constant (for  $M \approx N$). Finally, the bound increases for $N>M$ (Fig.~\ref{bipartite_2D5}).

\begin{figure}[h!tb]
\centering
\subfloat[]{
\includegraphics[trim = 15mm 69mm 25mm 70mm,clip,width=0.24\textwidth]{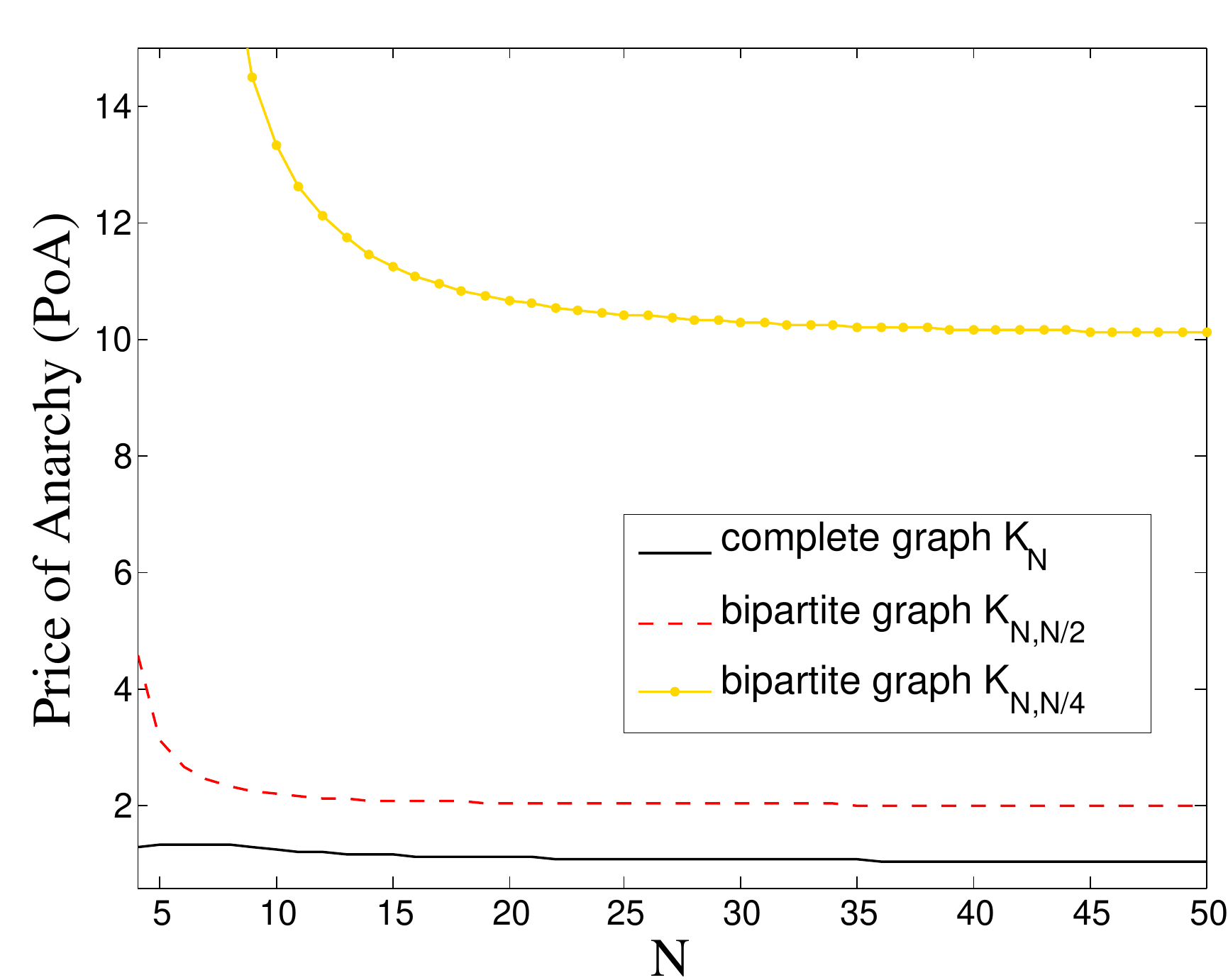}
\label{comparisonCompleteBipartite}}
\subfloat[]{
\includegraphics[trim = 15mm 69mm 25mm 70mm, clip,width=0.24\textwidth]{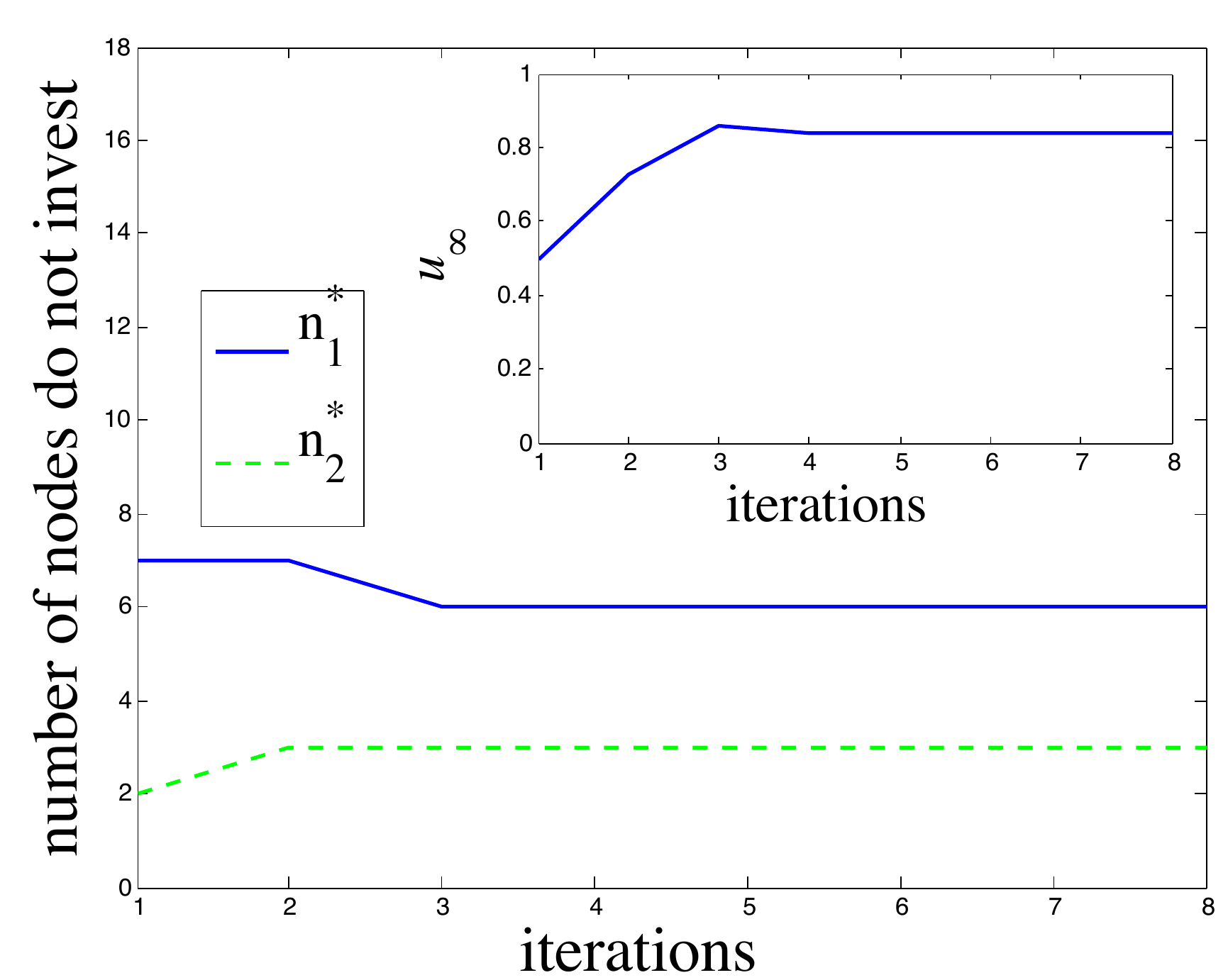}
\label{iter1anditerCN}}
\caption[]{(a) A comparison between PoA of a complete and bipartite graphs ($C=0.4$, $H=0.5$, $\tau = \frac{2}{3}$). (b) Main plot: the number of nodes that do not invest at the equilibrium. Inset: The infection probability of the core node $u_{\infty}$.} \label{fig:comparison}
\end{figure}

As shown in Fig.~\ref{comparisonCompleteBipartite}, the PoA in any bipartite graph is always higher than the one of a complete graph, if the same number of nodes are considered. In addition, PoA would have smaller value if the partitions have the similar sizes than having partitions with different sizes (e.g., a star graph).

%  More precisely, the lower bound of PoA increases in $N$ on the interval $[1,M]$, while it decreases in $N$ on the interval $[M,\infty)$ (in Fig.~\ref{bipartite_2D}). 

%\begin{figure}[h!]
%\begin{center}
%\includegraphics[trim = 0mm 0mm 0mm 0mm, clip,width=0.31\textwidth]{images/bpoabi.pdf}
%\end{center}
%\caption{Upper bound for the Price of Anarchy in bi-partite network as a function of $M$ and $N$.}
%\label{PoAUpperbound_in_M_and_N}
%\end{figure}

\subsection{Multi-communities network}

We illustrate the results obtained in Section~\ref{sec:CommunityNetwork}. Particularly, we present how the iterative algorithm can be used to obtain the equilibrium of the game by using the parametric potential game approach. First, we consider an example with $M=2$ communities with $N_1=10$, $N_2=15$, $\tau_1=0.5$ and $\tau_2=1.5$. Second, we consider the following stopping criteria $\varepsilon=10^{-7}$ and we observe that the number of iterations to achieve an equilibrium is very small (8 iterations).
In Fig.~\ref{iter1anditerCN} (main plot), we observe that the number of nodes which do not invest at equilibrium is $n_1^*=6$ (over 10 nodes) and $n_2^*=3$ (over 15 nodes). The iterative procedure starts with an initial probability of the core node $u_{\infty}[0]=0.5$. The infection probability of the core node obtained at the convergence of the iterative procedure is $u_{\infty}=0.8389$ (Fig.~\ref{iter1anditerCN} inset). We also consider a more complex scenario with $M=7$ communities and the following features $(N)=(10,15, 12, 8, 9, 4, 15)$ and $(\tau)=(0.5, 1.5, 1, 1.2, 1.4, 0.8, 0.1)$. The iterative algorithm converges after only 10 iterations for a precision $\varepsilon=10^{-7}$ and $u_{\infty}[0]=0.5$. We obtain the infection probability of the core node $u_{\infty}=0.935$ and the following equilibrium $n^*=(6,3,4,3,3,4,15)$. For the last community, for which the effective spreading rate is the lowest, none of the nodes invests at equilibrium, whereas, for the second community, for which the effective spreading rate is the highest, the number of nodes that invest is $12$ over $15$ nodes.
%\begin{figure}[h!]
%\begin{center}
%\includegraphics[trim = 32mm 95mm 40mm 100mm, clip,width=0.28\textwidth]{images/iter1_andCN.pdf}
%\end{center}
%\caption{The number of nodes that do not invest at the equilibrium in the main plot, while inset figure shows the infection probability of the core node $v_{\infty}$.}
%\label{iter1anditerCN}
%\end{figure}

\section{Conclusions}\label{sec:Conclusion}% and Perspectives
In this paper, we explore the problem of optimal decentralized protection strategies in a network, where a node decides to invest in an anti-virus or to be prone to the virus SIS epidemic spread process. If a node decides to invest, it cannot be infected, while if a node chooses not to invest, it can be infected by a virus and further spreads the virus inside the network. We study this problem from a game theoretic perspective. If a node decides to invest, the cost function of the node is the investment cost, otherwise the cost function is linearly proportional to the node's infection probability in the epidemic steady state.
%, expressed by epidemic rate $\beta$ or can used a self protection (e.g., a anti-spy and system cleaning tool to revive it self), expressed by the curing rate $\delta$.

We look for the optimal, fully decentralized protection strategies for each node. We show the existence of a potential structure, which allows us to prove the existence and uniqueness and derive the pure and mixed equilibrium in a \emph{single-community} (or \emph{mesh}) network. Moreover, we find the pure equilibrium in a \emph{bipartite} network. We also evaluate the performance of the equilibrium by finding the Price of Anarchy (PoA). Finally, we propose a simple, fully decentralized algorithm which converges to the pure equilibrium. In a \emph{multi-communities} topology, in which several communities are connected through a single core node, we introduce the concept of \emph{parametric potential games} and further derive an accurate, iterative heuristic for computing the Nash equilibrium.

\section*{Acknowledgment}
This research has been supported by the EU CONGAS project (project no. 288021). %S. T. would like to thank Ilina Kachinske (University of Maryland, College Park, USA) for the proofreading on the initial version.

\linespread{0.95}
%\linespread{1.0}

%\onecolumn
%\renewcommand\appendixname{Supplementary material}
\appendix{}

%\appendix{{\color{blue} [Given in this form for review purposes. In the final form, the appendix might evolve into a supplementary material, according to the TCNS rules and page limit.]}}
\section{}

\subsection{Proofs of the propositions and corollaries in a single community network}\label{ProofsCompleteGraph}

\subsubsection*{\textbf{Proposition~\ref{prep:equilibriumBounds}}}
\ifthenelse{\boolean{Longversion}}{
Let $S$ be the set of nodes that
invest $C$. A node $k\notin S$ must pay an infection cost equal to $H v_{\infty
}(n^*)$. If that node $k$ deviates\footnote{Deviation
means that the node changes its action. If the multistrategy for the node is to invest,
then \textquotedblleft deviating\textquotedblright\ means that the node does not
invest or \emph{vice versa}.}, then it will pay instead $C$. At equilibrium this user has no interest to deviate, meaning that: $
C \geq H v_{\infty} (n^*)
$.

It remains to show that nodes in $S$ do not
gain by deviating. Each node in $S$ pays constant cost $C$. When deviating, a node $l$,
originally in $S$, becomes connected to those not in $S$, which implies that
node $l$ changes the size of the set $N \setminus S$ which becomes $n^*+1$. The following inequality at equilibrium also applies: $
C \leq H v_{\infty} (n^*+1)
$.

Now, we show that $n^*$ exists and is uniquely defined. For $C>H$, we have a trivial solution $n^* =N$, otherwise based on (\ref{infection_prob_node_i_regular_graph}), we arrive at% $H (1-\frac{1}{(n^{*}-1)\tau}) \le C \le H (1-\frac{1}{n^{*} \tau})$
\small{
\begin{align*}
&H (1-\frac{1}{(n^{*}-1)\tau}) < C \le H (1-\frac{1}{n^{*} \tau})
\end{align*}
}\normalsize which gives $n^* = \lceil \frac{1}{(1-\frac{C}{H})\tau} \rceil $, %\frac{1}{(1- \frac{C}{H})\tau} \le n^* \le 1+ \frac{1}{(1- \frac{C}{H})\tau} \Leftrightarrow 
%\small{
%\begin{align*}
%\frac{1}{(1- \frac{C}{H})\tau} \le n^* \le 1+ \frac{1}{(1- \frac{C}{H})\tau} \Leftrightarrow n^* = \lceil \frac{1}{(1-\frac{C}{H})\tau} \rceil ,	
%\end{align*}
%}\normalsize 
if $\lceil \frac{1}{(1-\frac{C}{H})\tau} \rceil < N$, otherwise we have the trivial upper bound of $n^* =N$. \hfill $\blacksquare$
}
{The proof could be found in our technical Report~\cite{}.}

\subsubsection*{\textbf{Proposition~\ref{prep:SocOptComGraph}}}

\ifthenelse{\boolean{Longversion}}{
The cost of the nodes that do not invest is non-negative since $H \geq 0$ and $v_{\infty}(n) \ge 0$. If $n<1+\frac{1}{\tau}$,  $S(n)=C(N-n)$ which decreases in $n$. On the other hand, if $n \geq 1+\frac{1}{\tau}$, the derivative of (\ref{SW_pure}) is $
S'(n)=H-C+\frac{H}{\tau (n-1)^2}
$. Two cases can be distinguished:\\
%\begin{itemize}
1) $C<H$: the function $S(n)$ is strictly increasing over the interval $[1+\frac{1}{\tau},N]$, so the minimum is achieved in $\lceil 1+ \frac{1}{\tau} \rceil$.\\
2) $C \geq H$: the function $S(n)$ is increasing over the interval $[1+\frac{1}{\tau},1+\sqrt{\frac{H}{\tau(C-H)}}]$ and decreasing over $[1+\sqrt{\frac{H}{\tau(C-H)}},N]$, so the minimum is achieved in $\{\lceil 1+ \frac{1}{\tau} \rceil,N\}$ depending on the parameters of the system. \hfill $\blacksquare$
%\end{itemize}
}
{The proof could be found in our technical Report~\cite{}.}

\subsubsection*{\textbf{Corollary~\ref{optimimEquilibrium}}}

\ifthenelse{\boolean{Longversion}}{
If  $C \geq H$, based on the proof in Proposition~\ref{prep:equilibriumBounds}, we have $n^{*} = N$, which is clearly as large as any value of $n^{opt} \le N$. Otherwise ($C<H$), based on Proposition~\ref{PoABoundComp}: $ n^{opt} = \lceil 1+ \frac{1}{\tau} \rceil $ and $S(n)$ is increasing. Using the definition of PoA: $S(n^{opt})\leq S(n^*)$, which gives $n^{opt} \le n^*$. \hfill $\blacksquare$
}
{The proof could be found in our technical Report~\cite{}.}

\subsubsection*{\textbf{Corollary~\ref{PoABoundComp}}}

\ifthenelse{\boolean{Longversion}}{
First, the numerator $S(n^*)$ of $PoA_p$ is strictly lower than $CN$. Indeed, using Proposition~\ref{prep:equilibriumBounds} into (\ref{SW_pure}) we have: $S(n^*)=C(N-n^*)+n^*Hv_{\infty} (n^*) = CN - n^* (C- Hv_{\infty} (n^*)) < CN$.

In Proposition~\ref{prep:SocOptComGraph}, we obtain $n^{opt} \in \{N,\lceil 1+ \frac{1}{\tau}\rceil\}$. If $n^{opt} = N$ then $n^* = N$ (Corollary~\ref{optimimEquilibrium}) and $PoA_p=1$. For the case $n^{opt} =\lceil 1+ \frac{1}{\tau}\rceil $, the following (based on Proposition~\ref{prep:SocOptComGraph}) applies: (i) $C<H$: the function $S(n)$ is strictly increasing and\\ (ii) $C \geq H$ and $n^{opt} =\lceil 1+ \frac{1}{\tau}\rceil$, $S(n)$ is also strictly increasing. Therefore, in both cases $n^{opt} = \lceil 1+ \frac{1}{\tau} \rceil \ge 1+ \frac{1}{\tau}$, hence $S(n^{opt}) \ge S(1+ \frac{1}{\tau}) = C(N-(1+ \frac{1}{\tau}))$ i.e. $\text{$PoA_p$}=\frac{S(n^*)}{S(n^{opt})} \le \frac{CN}{C(N-(1+ \frac{1}{\tau}))} = \frac{1}{1-(1+\frac{1}{\tau})\frac{1}{N}}$. \hfill $\blacksquare$
%\small{
%\begin{align*}
%\text{$PoA_p$}&=\frac{SW(n^{opt})}{SW(n^*)} \ge 1-(1+\frac{1}{\tau})\frac{1}{N}.
%\end{align*}}
}{The proof could be found in our technical Report~\cite{}.}

\subsubsection*{\textbf{Proposition~\ref{prep:ExistUniqComp}}}

\ifthenelse{\boolean{Longversion}}{
For any $p \in [0,1]$ and any player $i$, we have: $
\bar{S}_{i}(1,p)=C\sum_{n=0}^{N-1}\binom{N-1}{n}(1-p)^np^{N-1-n}=C
$. We also have: $
\bar{S}_{i}(0,0)=H(1-\frac{1}{(N-1)\tau})>0
$, and $\bar{S}_{i}(0,1)=0$.

If $C<H(1-\frac{1}{(N-1)\tau})$ the mixed strategy $p^*$ exists because the function $\bar{S}_{i}(0,p)$ is continuous. Otherwise, we have for all $p \in [0,1]$, $\bar{S}_{i}(1,p)>\bar{S}_{i}(0,p)$, meaning that the strategy 0 is dominant irrespective of the mixed strategy of the other players. In this case, the action 0 is the equilibrium. \hfill $\blacksquare$
}
{The proof could be found in our technical Report~\cite{}.}

\subsubsection*{\textbf{Proposition~\ref{prep:MixedUniqEquil}}}

\ifthenelse{\boolean{Longversion}}{
The proof relies on the monotonicity of $\bar{S}_{i}(0,p)$ and the fact that $\bar{S}_{i}(1,p) = C$ (a horizontal line), so the two curves intersect in one point. However, proving the monotonicity of $\bar{S}_{i}(0,p)$ is not trivial, because the well know function $H\sum_{n=\underline{n}}^{N-1}\binom{N-1}{n} (1-p)^np^{N-1-n}$ decreases faster than $\bar{S}_{i}(0,p)$ on some intervals of $p$, but slower on other intervals. In what follows, we prove the monotonicity of $\bar{S}_{i}(0,p)$. For simplicity, we denote $M_{n}^{N}(p) = (1-\frac{1}{\tau n}) \binom{N-1}{n} (1-p)^n p^{N-1-n}$. Taking the first derivative in $p$ and using the fact that $1-\frac{1}{\tau n}<1-\frac{1}{\tau (n+1)}$, we obtain
\small{
\begin{align}
&\frac{d(M_{n}^{N}(p))}{dp} = (N-1)  (1-\frac{1}{\tau n}) \left[C_{n}^{N-1} (1-p) - C_{n-1}^{N-1} (1-p) \right] \nonumber\\ 
&\hspace{-0.5em}< (N-1) \big[ (1-\frac{1}{\tau (n+1)}) C_{n}^{N-1} (1-p) - (1-\frac{1}{\tau n}) C_{n-1}^{N-1} (1-p) \big]% \notag\\
%& = (N-1) \big(M_{n}^{N-1}(p) - M_{n-1}^{N-1}(p)\big)
 \label{monotonicityTemp}
\end{align}
}\normalsize where $C_{n}^{N-1} (1-p) = \binom{N-1}{n} (1-p)^np^{N-1-n}$ is \emph{Bernstein Basis Polynomial}. Summing (\ref{monotonicityTemp}) over all $n = \bar{n},\ldots, N-1$, multiplied by $H$ and knowing the values of the boundary term $C_{N-1}^{N-1} (1-p) = \binom{N-2}{N-1} (1-p)^{N-1} p^{-1} = 0$ result with%~\cite{lorentz1953bernstein}
\footnotesize{
\begin{align*}
&\frac{d(\bar{S}_{i}(0,p))}{dp} = H \sum_{n=\bar{n}}^{N-1} \frac{d(M_{n}^{N}(p))}{dp} < (N-1) H \times \\
&\big [ \sum_{n=\bar{n}}^{N-1} (1-\frac{1}{\tau (n+1)}) C_{n}^{N-1} (1-p) - \sum_{n=\bar{n}-1}^{N-2} (1-\frac{1}{\tau (n+1)}) C_{n}^{N-1} (1-p) \big]\\
&= (N-1) H \Big [(1-\frac{1}{\tau N}) C_{N-1}^{N-1} (1-p) - (1-\frac{1}{\tau \bar{n}}) C_{\bar{n}-1}^{N-1} (1-p) \Big ] = \\
&- (N-1) H (1-\frac{1}{\tau \bar{n}}) C_{\bar{n}-1}^{N-1} (1-p) < 0
\end{align*}
}\normalsize i.e. $\bar{S}_{i}(0,p)$ is a decreasing function.  \hfill $\blacksquare$
}
{The proof could be found in our technical Report~\cite{}.}

\subsubsection*{\textbf{Proposition~\ref{prep:SymmetricMixedComp}}}

\ifthenelse{\boolean{Longversion}}{
Using (\ref{infection_prob_node_i_regular_graph}), the average approximated cost is: $\hat{S}_i(p_i,p')= p_iC+(1-p_i)Hv_{i\infty}(\bar{n}+1) = p_iC+\\(1-p_i)H\max\{0,1-\frac{1}{\tau (1-p') (N-1)}\}$.
%\begin{align*}
%\hat{S}_i(p_i,p')=&p_iC+(1-p_i)Hv_{i\infty}(\bar{n}+1)\\
% %=& p_iC+(1-p_i)H(1-\frac{1}{\tau \bar{n}})\\
% =& p_iC+(1-p_i)H(1-\frac{1}{\tau (1-p) (N-1)})
%\end{align*}
Based on Definition~\ref{def:equilibrium} for the equilibrium, assuming that is achieved for $p'=\hat{p}^{*}$, we have $\hat{S}_i(0,\hat{p}^{*})=\hat{S}_i(1,\hat{p}^{*})$, which gives $C= H(1-\frac{1}{\tau (1-p^{*}) (N -1)})$ for $C < H(1-\frac{1}{\tau (N-1)})$ and $p^{*} C = 0$ otherwise. Finally, $\hat{p}^{*} = 1-\frac{H}{\tau (H-C)(N-1)}$, if $C < H(1-\frac{1}{\tau (N-1)})$, otherwise $\hat{p}^{*} = 0$ is an equilibrium.%, according to Proposition~\ref{prep:ExistUniqComp}.  %\newline \phantom{1}
 \hfill $\blacksquare$
%\begin{align*}
%\hat{S}_i(0,\hat{p}^{*})=&\hat{S}_i(1,\hat{p}^{*}) \Leftrightarrow C=   H(1-\frac{1}{\tau (1-p^{*}) (N -1)}) \\
%\hat{p}^{*} =& 1-\frac{H}{\tau (H-C)(N-1)}%1-\frac{H}{\tau (N-1)(H-C)}
%\end{align*}
}
{The proof could be found in our technical Report~\cite{}.}

\subsubsection*{\textbf{Proposition~\ref{prep:optimalAverage}}}

\ifthenelse{\boolean{Longversion}}{
First, the function is continuous in $p$, because the value is the same from the left and the right side of $1-\frac{1}{\tau(N-1)}$.  If $p\in[ 1-\frac{1}{\tau(N-1)},1]$ then it is increasing for any $C$ and $H$. If $C>H$, the function is increasing on the whole interval $p \in [0,1]$, hence $\hat{p}^{opt}=0$ and the value is $H(1-\frac{1}{\tau(N-1)})$. If $C=H$ the function is constant on $[0,1-\frac{1}{\tau(N-1)}]$, hence $\hat{p}^{opt} \in [0,1-\frac{1}{\tau(N-1)}]$ and the value is $H(1-\frac{1}{\tau(N-1)})$. If $C<H$ the function is decreasing on $[0,1-\frac{1}{\tau(N-1)}]$, hence $\hat{p}^{opt}=1-\frac{1}{\tau(N-1)}$ and the value is $C(1-\frac{1}{\tau(N-1)})$. \hfill $\blacksquare$
}
{The proof could be found in our technical Report~\cite{}.}

\vspace{1em}
\subsection{Proofs of the propositions and corollaries in a bipartite network}\label{ProofsBiPartite}

\subsubsection*{\textbf{Proposition~\ref{prep:equilibriumBoundsBipartite}}}\label{prep:equilibriumBoundsBipartiteProofApp}
For simplicity, we define $q = \frac{C}{H}$. By definition, at equilibrium a user from each of the two partitions has no interest to change its pure strategy, i.e. a player that decides to invest has no interest to change its strategy. Let $S$ be the set of nodes that
invest $C$. A node $k\notin S$ must pay an infection cost equal to $H \cdot v_{\infty}^{(\mathcal{M})}(n^*,m^*)$. If that node $k$ deviates, then it will pay instead $C$. At equilibrium a user from each of the partitions has no interest to deviate, meaning that: $C \geq H \cdot v_{\infty}^{(\mathcal{M})}(n^*,m^*)$ and $C \geq H \cdot v_{\infty}^{(\mathcal{N})}(n^*,m^*)$.

It remains to show that nodes in $S$ do not
gain by deviating. These nodes pay $C$ each. When deviating, a node $l$ in partition $\mathcal{N}$ nodes,
originally in $S$ becomes connected to those not in $S$, which implies that
node $l$ changes the size of $N \setminus S$ which becomes $n^*+1$. The following inequality at equilibrium also applies: $C \leq H \cdot v_{\infty}^{(\mathcal{N})}(n^*+1,m^*)$. In a similar way, considering a node in partition $\mathcal{M}$ nodes, it also holds $C \leq H \cdot v_{\infty}^{(\mathcal{M})}(n^*,m^*+1)$.

If $q = \frac{C}{H} > 1$ then for $m^*<M$ or $n^*<N$, we obtain a contradiction in the relations in the previous paragraph as $v_{\infty}^{(\mathcal{M})}(n^*,m^*+1)$ or $v_{\infty}^{(\mathcal{N})}(n^*+1,m^*)$ will be greater than $1$. Hence, the only possible value is $(m^*,n^*) = (M,N)$. Based on the discussion in the previous paragraph and the exact expression in (\ref{solutionsBipartite}) we end up with the following
\vspace{-0.6em}
\small{
\begin{align}
m<\frac{1}{\tau (\tau n (1-q)-q)} \le m+1 \label{firstBipartite}\\
n<\frac{1}{\tau (\tau m (1-q)-q)} \le n+1 \label{secondBipartite}
\end{align}
}\normalsize Hence, we have $m = \lceil \frac{1}{\tau (\tau n (1-q)-q)} \rceil -1 $ and $n = \lceil \frac{1}{\tau (\tau m (1-q)-q)} \rceil -1$, from which for a given $n$, we have a unique $m$ or \emph{vice versa}, which proves point 1) of the proposition.

Let us assume that $q \ge \frac{1}{2}$ or $\tau \ge \frac{(1+q)(1-2q)}{2q(1-q)}$. For simplicity in the derivations we denote $A = \tau^2 (1-q)$ and $B = \tau q$. From (\ref{secondBipartite}), we get $\frac{1}{Am-B}-1 \le n$, hence $\frac{A}{Am - B}- A - B \le An-B <\frac{1}{m}$. From the last inequality, we obtain $\frac{B}{A+B} \le m (Am-B)$. Finally,
\vspace{-1em}
\small{
\begin{align}
m > \frac{B+\sqrt{B^2+\frac{4AB}{A+B}}}{2A} \label{m_lower_bound}
\end{align}
}\normalsize Further, from (\ref{secondBipartite}) we obtain $Am-B <\frac{1}{n}$, hence $\frac{A(B+\sqrt{B^2+\frac{4AB}{A+B}})}{2A} -B \le Am-B <\frac{1}{n}$ i.e. $\frac{\sqrt{B^2+\frac{4AB}{A+B}}-B}{2} < \frac{1}{n} \Leftrightarrow n < \frac{2}{\sqrt{B^2+\frac{4AB}{A+B}}-B} = \frac{2 (B+\sqrt{B^2+\frac{4AB}{A+B}})}{\frac{4AB}{A+B}} = \frac{(1+\frac{A}{B}) (B+\sqrt{B^2+\frac{4AB}{A+B}})}{2A}$ i.e.
\vspace{-1em}
\small{
\begin{align}
n < \frac{(1+\frac{A}{B}) (B+\sqrt{B^2+\frac{4AB}{A+B}})}{2A} \label{n_upper_bound}
\end{align}
}\normalsize
From (\ref{m_lower_bound}) and (\ref{n_upper_bound}), we arrive at
\small{
\begin{align}
n-m &< \frac{A}{B}\frac{B+\sqrt{B^2+\frac{4AB}{A+B}}}{2A} = \frac{1+\sqrt{1+\frac{4A}{B(A+B)}}}{2} \nonumber\\ 
&=\frac{1+\sqrt{1+\frac{4\tau^2 (1-q)}{\tau^2 q(\tau (1-q)+q)}}}{2} = \frac{1+\sqrt{1+\frac{4 (1-q)}{q(\tau (1-q)+q)}}}{2} \label{temp_ineq}
\end{align}
}\normalsize The condition $q \ge \frac{1}{2}$ or $\tau \ge \frac{(1+q)(1-2q)}{2q(1-q)}$ is equivalent to $\frac{1-q}{k(\tau (1-q)+q)}<2$, and applying this in  (\ref{temp_ineq}), gives $n-m < \frac{1+\sqrt{1+4\cdot 2}}{2} = 2$. In the same way, $m-n <2$, hence $|n-m| \le 1$, which completes the proof for 2). In conclusion, we find out to have limited number of possibilities to be checked $n=m$;  $n=m-1$ or $n=m+1$ , from which the system of (\ref{firstBipartite}) and (\ref{secondBipartite}) boils down to significantly simplified one in one variable.

For 3), we give a counter example. We set $\tau = \frac{1+\sqrt{40 000 001}}{20 000} \approx 0.316278$ and $k = \frac{\sqrt{40 000 001}-1}{20 000}\approx 0.000316178$. Now, $A = 10^{-1}$ and $B = 10^{-4}$. The system of equations (\ref{firstBipartite}) and (\ref{secondBipartite}) will give $6$ solutions: $(n^*,m^*) = \{(1,10),(2,5),(3,3),(5,2),(10,1)\}$. Five of these solutions are pairs of numbers that are neither equal nor consecutive integers. \vspace{0em}\hfill $\blacksquare$%, which is not covered by the conditions in 2)

\subsubsection*{\textbf{Proposition~\ref{prep:SocOptBipartite}}}
%: (i) $n=N$ and $m = \frac{1}{\tau^2 N}$ or (ii)  Now, $SW\ge N+M-n-\frac{1}{\tau^2 n}$
$S(n,m)$ is a function of two variables. Bellow the epidemic threshold ($mn \le \frac{1}{\tau^2}$) and $S(n,m) = C (N+M-n-m)$.  If $M N \le \frac{1}{\tau^2}$, then $(n^{opt},m^{opt}) = (N,M)$ is the optimal pair and $S = 0$. In the remaining cases ($M N > \frac{1}{\tau^2}$), because the first derivatives in both $m$ and $n$ give constant non-zero values, we look for the extremal points on the boundaries in $m,n$-plane: $mn = \frac{1}{\tau^2}$; $m=M$ or $n=N$.\\
1) If $m=M$, then $n = \min \{N,\frac{1}{\tau^2 M}\} = \frac{1}{\tau^2 M}$ and $S =C \frac{\tau^2 MN-1}{\tau^2 M}$.\\
2) Similarly, if $n=N$, then $m = \frac{1}{\tau^2 N}$ and $S = C\frac{\tau^2 M N-1}{\tau^2 N}$.\\
3) If $mn= \frac{1}{\tau^2}$ then $\frac{1}{\tau^2 M} \le n \le N$ and $S(n) = C(M+N-n - \frac{1}{\tau^2 n})$. $S(n)$ increases to some point ($n = \frac{1}{\tau}$) and then starts to decrease, hence the minimum is on one of the boundaries, in the same points as 1) and 2). Finally, we take the minimum of 1) and 2), which gives $S = C\frac{\tau^2 MN-1}{\tau^2 \max \{ M,N\}}$, achieved for $(n^{opt},m^{opt}) = (\frac{1}{\tau^2 M},M)$  for $M>N$; $(n^{opt},m^{opt}) = (N,\frac{1}{\tau^2 N})$  for $M<N$ or both points for $M=N$.

Above the epidemic threshold ($mn \ge \frac{1}{\tau^2}$), we have
\vspace{-1em}
\small{
\begin{align*}
S(n,m) = C(N+M-n-m) + H(\frac{\tau^2 m n - 1}{\tau (\tau n + 1)}+\frac{\tau^2 m n - 1}{\tau (\tau m + 1)}) 
\end{align*}
}\normalsize Taking the first derivatives and equaling them to $0$, results with
\vspace{-1em}
\small{
\begin{align}%
S'_{n}(n,m) &= - C + H (\frac{\tau m }{\tau m+1} +\frac{\tau m +1}{(\tau n+1)^2}) = 0 \notag \\
S'_{m}(n,m) &= - C + H (\frac{\tau n }{\tau n+1} +\frac{\tau n +1}{(\tau m+1)^2}) = 0 \label{systemSingularities}
\end{align}
}\normalsize Subtracting the two equations of (\ref{systemSingularities}) gives
\small{
\begin{align*}
\frac{H \tau (2+ (m + n)\tau)^2 (m-n)}{(1+ \tau m)^2(1+ \tau n)^2}&=0
\end{align*}
}\normalsize Therefore $m=n$ is the only possibility. Going back into the first equation of (\ref{systemSingularities}) results with $C = H$. Hence, if $C\neq H$ there is no singular point inside the region and we should again look for the extrema on the boundaries: $mn\ge \frac{1}{\tau^2}$, $n\le N$ and $m \le M$.\\
1) if $mn = \frac{1}{\tau^2}$, then $S(n,m) =  C(N+M-n-m)$, so we again end up with the same solution as for the case bellow the epidemic threshold, considered before.\\
2) if $m=M$, we have $S(n) = C(N-n)+H(\frac{\tau^2 M n - 1}{\tau (\tau n + 1)}+\frac{\tau^2 M n - 1}{\tau (\tau M + 1)})$. The first derivative is $S'(n) = - C + H (\frac{\tau M }{\tau M+1} +\frac{\tau M +1}{(\tau n+1)^2}) $ and $S''(n) = -H\frac{\tau(\tau M +1)}{(\tau n+1)^3} <0$. Therefore, the function could only have local maximum and we should look for the minimum on the extremal points for $\frac{1}{\tau^2 M}\le n \le N$.
\begin{itemize}
\item $n = \frac{1}{\tau^2 M}$ then $S = C(N-\frac{1}{\tau^2 M})$ for $(n^{opt},m^{opt}) = (\frac{1}{\tau^2 M},M)$, which is again a boundary case exactly on the epidemic threshold and it was considered above.

\item $n=N$ then $S = H (\frac{\tau^2 M N - 1}{\tau (\tau N + 1)}+\frac{\tau^2 M N - 1}{\tau (\tau M + 1)}) = H \frac{(\tau^2MN-1)[\tau(M+N)+2]}{\tau(\tau M+1)(\tau N+1)}$ for $(n^{opt},m^{opt}) = (N,M)$.
\end{itemize}
3) if $n=N$, we have similar cases as in 2).

If $C=H$ (i.e. $k=1$) and $m=n$, then the social cost function boils down to $S(n) = C(N+M-2n) + 2C \frac{\tau^2 n^2 - 1}{\tau (\tau n + 1)} = C(N+M-2n) + 2C \frac{\tau n - 1}{\tau} = C(N+M-\frac{2}{\tau}) =  \text{const}$  i.e. S is constant and does not depend on $n$ or $m$. However, $C(N+M-\frac{2}{\tau})\ge C\frac{\tau^2 MN-1}{\tau^2 M}$ for any $m$ as the last is equivalent to $(\sqrt{M}-\frac{1}{\tau \sqrt{M}})^2 \ge 0$. In conclusion, $S = \max\{\tau^2 MN-1,0\}\cdot\min \{\frac{C}{\tau^2 \max \{ M,N\}},H \frac{\tau(M+N)+2}{\tau(\tau M+1)(\tau N+1)}\}$.\phantom{1} \hfill $\blacksquare$
\vspace{0.6em}
\subsubsection*{\textbf{Corollary~\ref{PoABound}}}
%The PoA is defined by:
%$$
%PoA=\frac{SW(n^{opt})}{SW(n^*)}.
%$$
First, the denominator of PoA: $S(n^*,m^*)$ is strictly lower than $C(N+M)$. Indeed, using Proposition~\ref{prep:equilibriumBoundsBipartite} and equation (\ref{SW_pureBip}) we have:
\small{
\begin{align*}
S(n^*,m^*) =& C(N+M-n^*-m^*)\\
&+n^* H v_{\infty}^{(\mathcal{N})}(n^*,m^*) + m^*H v_{\infty}^{(\mathcal{M})}(n^*,m^*) \\
=& C(N+M) - n^* (C- Hv_{\infty}^{(\mathcal{N})}(n^*,m^*)) \\
&- m^* (C- Hv_{\infty}^{(\mathcal{M})}(n^*,m^*)) < C(N+M)
\end{align*}
}\normalsize which is an upper bound for $S(n^*,m^*)$ and $S(n^{opt},m^{opt})$ could be determined exactly based on Proposition~\ref{prep:SocOptBipartite}, which completes the proof. \hfill $\blacksquare$

\subsubsection*{\textbf{Corollary~\ref{PoALowerBound}}}

For simplicity, we denote $h(M,N,\tau) = \frac{\tau(M+N)}{\max\{\tau^2 MN-1,0\} \min \{\frac{1}{\tau \max \{ M,N\}},\frac{H(\tau(M+N)+2)}{C(\tau M+1)(\tau N+1)}\}}$. If $\tau^2 MN\le1$, then $h(M,N,\tau) = \infty \ge \max \{2,\frac{C}{H} \} $. If $\tau^2 MN>1$, then
\small{
\begin{align*}h(M,N,\tau) &\ge \frac{\tau^2 \max \{ M,N\} (M+N)}{\tau^2 MN-1} = \frac{\max \{ M,N\} (\frac{1}{M}+\frac{1}{N})}{1-\frac{1}{\tau^2 MN}} \\
&= \frac{1+\frac{\max \{ M,N\}}{\min \{ M,N\}}}{1-\frac{1}{\tau^2 MN}} > \frac{1+1}{1-0} = 2 \text{        \normalsize and}
\end{align*}
}
\vspace{-1em}
\small{
\begin{align*}&h(M,N,\tau) \ge \frac{C}{H}\frac{\tau (M+N)(\tau M+1)(\tau N+1)}{(\tau^2 MN-1)(\tau(M+N)+2)} \ge \frac{C}{H} \times\\
&  \frac{\tau^3 MN (M+N) + \tau^2 (2MN+M^2+N^2) + \tau (M+N)}{\tau^3 MN (M+N) + \tau^2 (2MN) - \tau (M+N) -2}> \frac{C}{H}.
\end{align*}
}\normalsize \hfill $\blacksquare$
\vspace{-1.05em}
\subsection{Technical details of the expressions for a multi-communities network}\label{ProofsMultiComm}

\subsubsection*{\textbf{Details of expression~(\ref{prep:ParamPotCommGame})}}

\ifthenelse{\boolean{Longversion}}{
For simplicity, we denote $V (\tau_{m},n_m, u_{\infty}) = \tau _{m}(n_{m}-1)-\tau _{m} u_{\infty}-1$ and $W (\tau_m,n_m, u_{\infty}) = \left( \tau _{m}(n_{m}-1)\right) ^{2}+\left( \tau _{m}u_{\infty }\right)
^{2}+1+2\left( \tau _{m}u_{\infty }\right) +2\tau_{m} u_{\infty} \tau _{m}(n_{m}-1)$.
For a given $u_{\infty }$, for any non-core node in community $\mathcal{N}_m$, based on equation (\ref{steady_state_NIMFA_node_i_1}), we have the equation 
\small{
\begin{align*}
\tau
_{m}(n_{m}-1) (v_{\infty}^{(\mathcal{N}_m)})^2 - V (\tau_{m},n_m, u_{\infty}) v_{\infty}^{(\mathcal{N}_m)} - \tau_m u_{\infty} = 0  
\end{align*}
}\normalsize of degree $2$ in $v_{\infty}^{(\mathcal{N}_m)}(n_m,u_{\infty }) $, which solutions are the following infection probabilities: $v_{\infty}^{(\mathcal{N}_m)}(n_m,u_{\infty }) 
 = \frac{V (\tau_{m},n_m, u_{\infty})( 1 \pm \sqrt{1+\frac{4\tau
_{m}^{2}u_{\infty }(n_{m}-1)}{V (\tau_{m},n_m, u_{\infty})}})}{2\tau _{m}(n_{m}-1)}=  \frac{V (\tau_{m},n_m, u_{\infty}) +\sqrt{W (\tau,n_m, u_{\infty}) -2\tau _{m}(n_{m}-1)}}{2\tau
_{m}(n_{m}-1)}$, where $V (\tau_{m},n_m, u_{\infty})$ is positive, otherwise, we would have two negative solutions. The value under the square root is greater than $1$, so the solution with ``-" sign is negative, hence it is not valid. Using the fact that $-2\tau _{m}(n_{m}-1)<2\tau _{m}(n_{m}-1)$, we obtain $v_{\infty}^{(\mathcal{N}_m)}(n_m,u_{\infty })  <\frac{V (\tau_{m},n_m, u_{\infty}) +\sqrt{W (\tau,n_m, u_{\infty}) + 2\tau _{m}(n_{m}-1)}}{2\tau
_{m}(n_{m}-1)} =1  $. Hence, the solution with sign ``$+$" is in the interval $\left( 0,1\right) $.% \hfill $\blacksquare$
}
{The proof could be found in our technical Report~\cite{}.}

\subsubsection*{\textbf{Bounds of $u_{\infty}$ and a discussion on the algorithm convergence}} For simplicity, we define $q = \frac{C}{H}$. Here, we show that $g(u_{\infty}[k]) <u_{\infty}[k+1]<f(u_{\infty}[k])$, where $f$ and $g$ are decreasing functions bounded from both sides. The functions $f$ and $g$ do not converge to the same value, hence an absolute convergence based on this result cannot be stated. In this direction, there might be extreme cases of non-convergence if $u_{\infty}$ changes from increasing to decreasing or \emph{vice versa}, periodically. However, this bounding is an evidence that $u_{\infty}$ converges (in practice).  Moreover, if $u_{\infty}$ is monotone then the convergence is proved.% and in the case of neglecting the ``integer nature (rounding)'' of $n_m^{*t}$.

Applying the condition for a Nash Equilibrium (or finding the minimum of the potential function) for the game in each community, gives $v_{\infty}^{(\mathcal{N}_m)}(n_m^{*}[k],u_{\infty }[k]) < \frac{C}{H} = q < v_{\infty}^{(\mathcal{N}_m)}(n_m^{*}[k]+1,u_{\infty }[k])$. Using the expression from Proposition~\ref{prep:ParamPotCommGame}, we obtain
\small{
\begin{align}
\frac{1}{\tau_m (1-q)} - \frac{u_{\infty}[k]}{q} < n_m^{*}[k] < \frac{1}{\tau_m (1-q)}+1 - \frac{u_{\infty}[k]}{q} , \label{boundNumberNodesMulti}
\end{align}
}\normalsize i.e. $n_m^{*}[k] = \lfloor\frac{1}{\tau_m (1-q)} - \frac{u_{\infty}[k]}{q} \rfloor$ for $q<1$, otherwise $n_m^{*}[k] = N_m$. We proceed with the case $q<1$. (For $q \ge 1$, a similar, but simpler analysis applies, because $n_m^{*}[k] = N_m$ is constant over $k$.) We continue with bounding $\tau_m n_m^{*}[k] v_{\infty}^{(\mathcal{N}_m)}(n_m^{*}[k],u_{\infty }[k]) $.

1) For the upper bound, using the right part of (\ref{boundNumberNodesMulti}), we have $\tau_m n_m^{*}[k] v_{\infty}^{(\mathcal{N}_m)}(n_m^{*}[k],u_{\infty }[k])  < q\tau_m n_m^{*}[k] < \frac{q}{1-q} + q \tau_m - \tau_m u_{\infty}[k]$.

2) For the lower bound, using the left part of (\ref{boundNumberNodesMulti}), we arrive at $\tau_m n_m^{*}[k] v_{\infty}^{(\mathcal{N}_m)}(n_m^{*}[k],u_{\infty }[k]) > \tau_m (n_m^{*}[k] -1) v_{\infty}^{(\mathcal{N}_m)} = \frac{\tau_m (n_m^{*}[k] -1) - \tau_m u_{\infty }[k] -1}{2} +\frac{ \sqrt{(\tau_m (n_m^{*}[k] -1) - \tau_m u_{\infty }[k] -1)^2 + 4 \tau_m^2 (n_m^{*}[k] -1)u_{\infty }[k] } }{2} > \frac{\frac{q}{1-q} - \frac{1+q}{q}u_{\infty }[k] \tau_m - \tau_m + \frac{q}{1-q} + \frac{1-q}{q}u_{\infty }[k] \tau_m - \tau_m}{2} = \frac{q}{1-q}  - \tau_m - \tau_m u_{\infty}[k]$.% (\ref{prep:ParamPotCommGame}) 

Now, applying 1) and 2) into the expression for $u_{\infty}[k+1]$ (see step 3) in the algorithm), yield
\footnotesize{
\begin{align}
&1- \frac{1}{1+\sum_{m=1}^{M} (\frac{q}{1-q} - \tau_m  - \tau_m u_{\infty}[k])} < u_{\infty}[k+1]  \nonumber\\
&=1-\frac{1}{1+\sum_{m=1}^{M}\tau_m n_m^{*}[k] v_{\infty}^{(\mathcal{N}_m)}(n_m^{*}[k],u_{\infty }[k])} \nonumber \\
&< 1- \frac{1}{1+\sum_{m=1}^{M} (\frac{q}{1-q} +q\tau_m - \tau_m u_{\infty}[k])} \text{\normalsize\hspace{1em}    or } \nonumber %\\
\end{align}
}\normalsize% or
\footnotesize{
\begin{align*}
&1- \frac{1}{(1+\frac{Mq}{1-q} -\sum_{m=1}^{M} \tau_m)  -  u_{\infty}[k] \sum_{m=1}^{M} \tau_m } < u_{\infty}[k+1] < \\
&1- \frac{1}{(1+\frac{Mq}{1-q} - \sum_{m=1}^{M} \tau_m) +(1+q) \sum_{m=1}^{M} \tau_m  -  u_{\infty}[k] \sum_{m=1}^{M} \tau_m }.
\end{align*}
}\normalsize Therefore, $u_{\infty}[k+1]$ is bounded from above and below, respectively, by two bounded decreasing functions $f(u_{\infty}[k]) = 1- \frac{1}{(1+\frac{Mq}{1-q} - \sum_{m=1}^{M} \tau_m) +(1+q) \sum_{m=1}^{M} \tau_m  -  u_{\infty}[k] \sum_{m=1}^{M} \tau_m }$ and $g(u_{\infty}[k]) = 1- \frac{1}{(1+\frac{Mq}{1-q} -\sum_{m=1}^{M} \tau_m)  -  u_{\infty}[k] \sum_{m=1}^{M} \tau_m } $. The function $f(u_{\infty}[k])$ contains the term $(1+q)\sum_{m=1}^{M} \tau_m$ in the denominator of the quotient, which is the only difference from $g(u_{\infty}[k])$. The Squeeze (Sandwich) theorem cannot be applied, because $f$ and $g$ converge to different values. However, if $u_{\infty}[k]$ is a monotone increasing sequence then, due to the fact that $f(u_{\infty}[k])$ is a decreasing function, $u_{\infty}[k]$ converges. Similarly, if $u_{\infty}[k]$ is a monotone decreasing sequence then, due to the fact that $g(u_{\infty}[k])$ is a decreasing function, $u_{\infty}[k]$  also converges. Both cases are possible, based on the initially taken $u_{\infty}[0]$, however $u_{\infty}[k]$ is not necessarily monotone sequence (see e.g., Fig.~\ref{iter1anditerCN}). However, in practice, $u_{\infty}[k]$ converges.

\linespread{0.99}

\linespread{0.9836}

\vspace{-0.7cm}
\begin{IEEEbiography}
[{\includegraphics[width=1in,height=1.25in,clip,keepaspectratio]{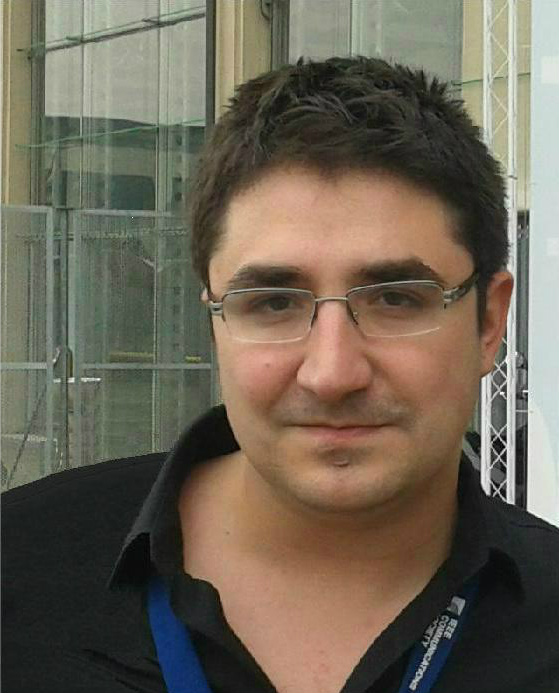}}]
%[{\includegraphics[width=1in,height=1.25in,clip,keepaspectratio]{bioImages/Stojan_bw.jpg}}]
{Stojan Trajanovski} received a PhD degree (\emph{cum laude}, 2014) from Delft University of Technology, The Netherlands. He obtained a master degree in Advanced Computer Science (\emph{with distinction}, 2011) from the University of Cambridge, United Kingdom. He also holds an MSc degree in Software Engineering (2010) and a Dipl. Engineering degree (\emph{summa cum laude}, 2008). He successfully participated in international science olympiads, winning a bronze medal at the International Mathematical Olympiad (IMO) in 2003.

His main research interests include network robustness, clustering analysis of networks and applied graph theory.
\end{IEEEbiography}

\vspace{-0.595cm}

\begin{IEEEbiography}
[{\includegraphics[width=1in,height=1.25in,clip,keepaspectratio]{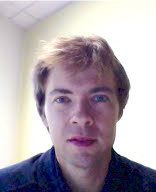}}]
%[{\includegraphics[width=1in,height=1.25in,clip,keepaspectratio]{bioImages/Yezekael_bw.jpg}}]
{Yezekael Hayel} received the M.Sc. degree in computer science and applied mathematics from the University
of Rennes 1, France, in 2002 and the Ph.D. degree in computer science from the University of Rennes 1 and INRIA in 2005. Since 2006, he has been an Assistant Professor at LIA/CERI, University of Avignon, France and since 2014, a visiting professor at NYU, USA for one year.. His research interests include performance evaluation of networks based on game theoretic and queuing
models. His research focuses on applications in communication networks, such as wireless flexible networks, bio-inspired and self-organizing networks, and economics models of the Internet. Since he has joined the networking group of the LIA/CERI, he has participated in several national (ANR) and international projects, such as the European and cefipra, with industrial companies like Orange Labs, Alcatel-Lucent, IBM, and academic partners like Sup\'{e}lec, CNRS, and UCLA. He has been invited to give seminal talks in institutions like INRIA, Sup\'{e}lec, UAM (Mexico), ALU (Shanghai).

\end{IEEEbiography}

\vspace{-0.595cm}

\begin{IEEEbiography}
[{\includegraphics[width=1in,height=1.25in,clip,keepaspectratio]{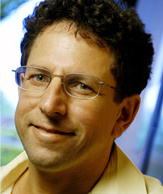}}]
%[{\includegraphics[width=1in,height=1.25in,clip,keepaspectratio]{bioImages/Eitan_bw.jpg}}]
{Eitan Altman} (M'93-SM'00-F'10) received the B.Sc. degree in electrical engineering, the B.A. degree in physics, and the Ph.D. degree in electrical engineering from the Technion - Israel Institute of Technology, Haifa, Israel, in 1984, 1984, and 1990, respectively, and the B.Mus. degree in music composition from Tel-Aviv University, Tel-Aviv, Israel, in 1990. Since 1990, he has been a Researcher with the National Research Institute in Computer Science and Control (INRIA), Sophia-Antipolis, France. He has been on the Editorial Boards of several journals: \emph{Wireless Networks}, \emph{Computer Networks}, \emph{Computer Communications}, \emph{Journal of Discrete Event Dynamic Systems}, \emph{SIAM Journal of Control and Optimisation}, \emph{Stochastic Models}, and \emph{Journal of Economy Dynamic and Control}. His areas of interest include networking, stochastic control, and game theory. Dr. Altman received the Best Paper Award in the Networking 2006, IEEE GLOBECOM2007, and IFIP Wireless Days 2009 conferences; and Best Student Paper awards at QoFis 2000 and at Networking 2002.
\end{IEEEbiography}

\vspace{-0.595cm}

\begin{IEEEbiography}
[{\includegraphics[width=1in,height=1.25in,clip,keepaspectratio]{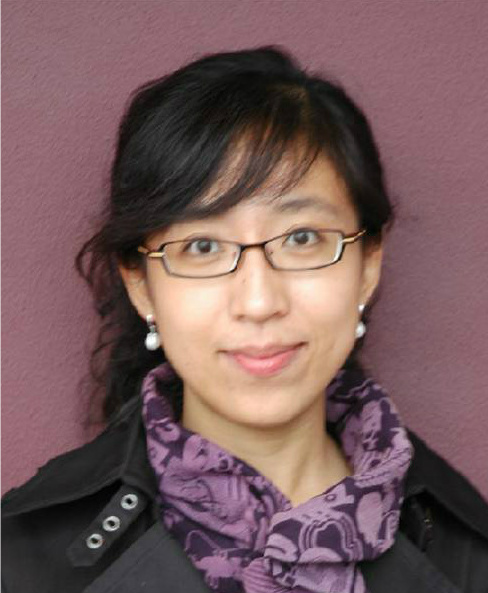}}]
%[{\includegraphics[width=1in,height=1.25in,clip,keepaspectratio]{bioImages/Huijuan_bw.jpg}}]
{Huijuan Wang} received her M.Sc. degree (\emph{cum laude}, 2005) in Electrical Engineering at the Delft University of Technology, the Netherlands. She obtained her Ph.D. degree (\emph{cum laude}, 2009) from the same university. She is currently an assistant professor in the Network Architecture and Services (NAS) Group at Delft University of Technology. She has been a visiting Professor at the group of Prof. H. Eugene Stanley at Boston University since August 2011. Her research interests are: multi-level, multi-scale complex networks and dynamic processes such as epidemic spread, opinion interactions and cascading failures, network structure design, and bio-inspired networking: from metabolic networks to brain networks.
\end{IEEEbiography}

\vspace{-0.595cm}

\begin{IEEEbiography}
[{\includegraphics[width=1in,height=1.25in,clip,keepaspectratio]{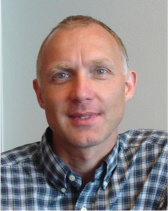}}]
%[{\includegraphics[width=1in,height=1.25in,clip,keepaspectratio]{bioImages/Piet_bw.jpg}}]
{Piet Van Mieghem}
received the Master’s (\emph{magna cum laude}, 1987) and PhD (\emph{summa cum laude}, 1991) degrees in electrical engineering from the K.U. Leuven, Leuven, Belgium. He is a Professor at the Delft University of Technology and Chairman of the section Network Architectures and Services (NAS) since 1998. His main research interests lie in modeling and analysis of complex networks and in new Internet-like architectures and algorithms for future communications networks. Before joining Delft, he worked at the Interuniversity Micro Electronic Center (IMEC) from 1987 to 1991. During 1993-1998, he was a member of the Alcatel Corporate Research Center in Antwerp, Belgium. He was a visiting scientist at MIT (1992-1993), a visiting professor at UCLA (2005) and Cornell University (2009); and will be a visiting professor at Stanford University (2015). He is the author of four books: Performance Analysis of Communications Networks and Systems (Cambridge
Univ. Press, 2006), Data Communications Networking (Techne, 2011), Graph Spectra for Complex Networks (Cambridge Univ. Press, 2011), and Performance Analysis of Complex Networks and Systems (Cambridge Univ. Press, 2014).

\end{IEEEbiography}

\end{document}